\documentclass[12pt,letterpaper]{article}%
\usepackage{amsmath,amssymb,amsfonts}
\usepackage{color}
\usepackage{float}
\usepackage{hyperref}
\usepackage[Symbolsmallscale]{upgreek}
\usepackage{amsmath}
\usepackage{amsfonts}
\usepackage{amssymb,dsfont}
\usepackage{graphicx}
\usepackage{amssymb}
\usepackage[vcentermath]{youngtab}
\usepackage[all]{xy}
\usepackage{pstricks}
\usepackage{dsfont}%
\begin{document}
\begin{titlepage}
\begin{flushright}
LPTENS-10/39
\end{flushright}
\vskip 1.0cm
\begin{center}
{\Large \bf Bounds in 4D Conformal Field Theories\\[5pt] with Global Symmetry}
\vskip 1.0cm
{\large Riccardo Rattazzi$^{a}$,\ Slava Rychkov$^{b}$,\ Alessandro Vichi$^a$} \\[0.7cm]
{\it $^a$ Institut de Th\'eorie des Ph\'enom\`enes Physiques, EPFL,  CH--1015 Lausanne, Switzerland\\[5mm]
$^b$ Laboratoire de Physique Th\'{e}orique, \'{E}cole Normale Sup\'{e}rieure,\\
and Facult\'{e} de Physique, Universit\'{e} Pierre et Marie Curie,
France}
\end{center}
\vskip 1.0cm
\begin{abstract}
We explore the constraining power of OPE associativity in 4D
Conformal Field Theory with a continuous global symmetry group. We
give a general analysis of crossing symmetry constraints in the
4-point function $\,\left\langle \phi
\phi\phi^{\dagger}\phi^{\dagger}\right\rangle $, where $\phi$ is a
primary scalar operator in a given representation $R$. These
constraints take the form of `vectorial sum rules' for conformal
blocks of operators whose representations appear in $R\otimes R$ and
$R\otimes\bar{R}$. The coefficients in these sum rules are related
to the Fierz transformation matrices for the  $R\otimes
R\otimes\bar{R}\otimes\bar{R}$ invariant tensors. We show that the
number of equations is always equal to the number of symmetry
channels to be constrained. We also analyze in detail two
cases---the fundamental of $SO(N)$ and the fundamental of $SU(N)$.
We derive the vectorial sum rules explicitly, and use them to study
the dimension of the lowest singlet scalar in the $\phi\times
\phi^{\dagger}$ OPE. We prove the existence of an upper bound on the
dimension of this scalar. The bound depends on the conformal
dimension of $\phi$ and approaches $2$ in the limit
$\dim(\phi)\rightarrow1.$ For several small groups, we compute the
behavior of the bound at $\dim(\phi)>1$. We discuss implications of
our bound for the Conformal Technicolor scenario of electroweak
symmetry breaking.
\end{abstract}
\vskip 1cm \hspace{0.7cm} September 2010
\end{titlepage}

\newpage

\section{Introduction}

Conformal Field Theory was originally conceived in four and three dimensions,
with applications to particle physics and critical phenomena in mind. However,
it is in 2D that the most spectacular results and exact solutions have been
obtained. In higher dimensions, there seems to be a general feeling that the
constraining power of conformal symmetry \textit{by itself} is insufficient to
tell nontrivial things about dynamics. Hence the interest in various
additional assumptions, like supersymmetry, or integrability for the planar
$\mathcal{N}=4$ super Yang-Mills, or the AdS/CFT duality. This is not fully
satisfactory, since there are likely many 4D CFTs which do not fulfill any of
these assumptions. For example, \textquotedblleft conformal windows" of
non-supersymmetric gauge theories.

And yet, in the early days of 4D CFT, it was hoped that the \textit{OPE
associativity} is such a strong constraint on the CFT data (the spectrum of
operator dimensions and the 3-point function coefficients) that it could allow
for a complete solution of the theory. Recently \cite{r1},\cite{r2}%
,\cite{cr},\cite{r3} we have been taking a fresh look at this idea, originally
proposed by Polyakov \cite{pol}. Our approach is not to try to \textit{solve}
the OPE associativity, but rather to try to deduce from it general bounds that
any CFT must obey. We discovered that such general bounds do exist! The bounds
found so far fall into two general classes:

\begin{itemize}
\item \cite{r1},\cite{r2} An upper bound on the gap in the operator spectrum:
\textit{any unitary 4D CFT containing a scalar operator }$\phi$\textit{ of
dimension }$d_{\phi}$\textit{ must also contain another scalar }$O$\textit{
appearing in the OPE }$\phi\times\phi$\textit{ whose dimension }$d_{O}%
$\textit{ is bounded by a universal function of }$d_{\phi}$\textit{:}%
\begin{equation}
d_{O}\leq f(d_{\phi}). \label{eq:bound}%
\end{equation}
The function $f(d_{\phi})$ has been computed numerically by means of a
well-defined algorithm.

\item Upper bounds on the 3-point function coefficients $\lambda_{\phi\phi O}$
where $\phi$ is the same scalar as above and $O$ is any primary operator in
the OPE $\phi\times\phi$. They have the form:%
\begin{equation}
|\lambda_{\phi\phi O}|\leq g(d_{\phi},d_{O},l_{O}), \label{eq:intr2}%
\end{equation}
where $g$ is a universal function of the dimensions and of the $O$'s spin
$l_{O}$. For $l_{O}=0$ this function was evaluated in \cite{cr}. Very
recently, the case $l_{O}=2,$ $d_{O}=4$, corresponding to the stress tensor
OPE coefficient, was considered in~\cite{poland},\cite{r3}. In this case the
upper bound (\ref{eq:intr2}) gives a lower bound on the central charge of the theory.
\end{itemize}

In this paper, we will discuss a generalization of bounds of the first class
to the case when CFT has a continuous global symmetry $G$ (Abelian or
non-Abelian), and an operator $\phi$ transforms in a nontrivial representation
$R$ of $G$. We will consider the OPE $\phi\times\phi^{\dagger}$ if $R$ is
complex, or $\phi\times\phi$ if $R$ is real. We will be discussing how to show
that this OPE necessarily contains a \textit{singlet} \textit{scalar }operator
$S$ whose dimension $d_{S}$ is bounded by a universal function which depends
only on $\phi$'s dimension and transformation properties:%
\begin{equation}
d_{S}\leq f_{S}(d_{\phi},R_{G}). \label{eq:boundS}%
\end{equation}
Thus the novelty with respect to \cite{r1},\cite{r2}\ is that we will be
bounding the gap in a given global symmetry sector (singlet in this case).

It is useful to recall that the original motivation of \cite{r1} was to find a
bound of precisely this type for the case $G=SO(4)$ and $\phi$ in the
fundamental. This in turn was needed in order to constrain the Conformal
Technicolor scenario of electroweak symmetry breaking \cite{luty}. This
connection was discussed extensively in \cite{r1}, and we will come back to it
in the discussion Section.

The paper is organized as follows. We first derive the \textquotedblleft
vectorial sum rules", which generalize the main equation encoding the OPE
associativity---the sum rule of \cite{r1}---to the globally symmetric case.
Not surprisingly, these new sum rules take a different form depending on $G$
and $R$ under consideration. To begin with, we treat three concrete examples:
$\phi$ in the fundamental of $SO(N)$; $\phi$ charged under a $U(1)$; $\phi$ in
the fundamental of $SU(N)$. Here we provide explicit derivations, illustrating
the necessary technical ingredients. We then consider the general case in some
detail, and in particular show that the number of constraints from crossing
symmetry is always equal to the number of unknown functions.

Then we discuss what our vectorial sum rules imply for the bounds on the
singlet dimension. In principle, functions $f_{S}$ can be computed by a
straightforward generalization of the algorithm of \cite{r1}. However,
numerical difficulties involved are much greater in the present case, because
one is working in a linear space of a much bigger dimension. As a result, we
cannot yet push our analysis to the point where it produces numerically
significant bounds. Thus we follow a more modest strategy. First, we explain
how one shows that a bound does exist for $d_{\phi}$ sufficiently close to 1.
The main idea is to consider the case $d_{\phi}=1$ and then to argue by
continuity (which is possible since the vectorial sum rules are continuous in
$d_{\phi}$). This analysis also shows that $f_{S}$ approaches $2$ as $d_{\phi
}\rightarrow1$. In other words, just like in our previous work, the free field
theory limit is approached continuously.

The next question is then to determine how fast $f_{S}$ approaches this limit.
Here we limit ourselves to quoting several $f_{S}$ values at $d_{\phi}>1$,
leaving the determination of a detailed shape of these bounds to future work.

We conclude by discussing consequences for phenomenology and promising
research directions.

\textbf{Note.} As this work was being prepared for publication, a very
interesting paper \cite{poland} appeared, which gave several important
generalizations of our method and results. In particular, \cite{poland}
generalized our method to the $\mathcal{N}=1$ SCFT case, and used it to derive
bounds on non-BPS quantities. They also derived lower bounds on the central
charge and on the two-point functions of global symmetry currents, with or
without supersymmetry. Finally, they presented a set of equations
incorporating OPE associativity constraints in the case of a $U(1)$ global
symmetry. They have even derived a bound on the $U(1)$ singlet dimension (in
the supersymmetric case). We will be commenting on this partial overlap in
more detail below.

\section{Sum rules in CFTs with a global symmetry}

\label{sect:sumrules}

\subsection{Conventions}

\label{sec:conventions}

We begin with some preliminary comments and notational conventions. We will
work in the Euclidean space. Just as in our previous work, availability of
explicit expressions for 4D conformal blocks given by Dolan and Osborn
\cite{do12} will play a crucial role. Consider a 4-point function
$\left\langle \phi(x_{1})\chi^{\dagger}(x_{2})\chi(x_{3})\phi^{\dagger}%
(x_{4})\right\rangle $ where $\phi$ and $\chi$ are two primary operators, not
necessarily Hermitean, assumed to have equal dimensions $d_{\phi}=d_{\chi}=d$.
The OPE $\phi\times\chi^{\dagger}$ will contain a sequence of spin $l,$
dimension $\Delta$ primary fields $O_{\Delta,l}$,:
\begin{equation}
\phi\times\chi^{\dagger}=\sum_{\Delta,l}\lambda_{\Delta,l}O_{\Delta,l}\,.
\end{equation}
Here $\lambda_{\Delta,l}$ are the OPE coefficients, in general complex. We
then normalize the conformal blocks via:%
\begin{gather}
\left\langle
\begin{array}
[c]{cc}%
\phi(x_{1})_{\bullet} & _{\bullet}\phi^{\dagger}(x_{4})\\
\chi^{\dagger}(x_{2})^{\bullet} & ^{\bullet}\chi(x_{3})
\end{array}
\right\rangle =\sum_{\Delta,l}\frac{1}{x_{12}^{2d}x_{34}^{2d}}\,p_{\Delta
,l}\,g_{\Delta,l}(u,v)\,,\label{eq:confblocks}\\
u\equiv x_{12}^{2}x_{34}^{2}/(x_{13}^{2}x_{24}^{2})=z\bar{z},\quad v\equiv
x_{14}^{2}x_{23}^{2}/(x_{13}^{2}x_{24}^{2})\,=(1-z)(1-\bar{z})\,,\\
g_{\Delta,l}(u,v)=+\frac{z\bar{z}}{z-\bar{z}}[k_{\Delta+l}(z)k_{\Delta
-l-2}(\bar{z})-(z\leftrightarrow\bar{z})]\,,\\
k_{\beta}(x)\equiv x^{\beta/2}{}_{2}F_{1}\left(  \beta/2,\beta/2,\beta
;x\right)  \,.
\end{gather}
The points $x_{i}$ are assumed to be near the vertices of a square, as the
picture suggests. The ordering is important. Eq.~(\ref{eq:confblocks}) says
that the exchanges of $O_{\Delta,l}$ and of its conformal descendants in the
(12)(34) channel ($\equiv$s-channel) can be summed up in a `conformal block'
$g_{\Delta,l}(u,v)$. The coefficients $p_{\Delta,l}$ are given by
\begin{equation}
p_{\Delta,l}=|\lambda_{\Delta,l}|^{2}>0\text{.}%
\end{equation}

Compared to \cite{do12}, and also to our previous work, we have dropped the
$(-1/2)^{l}$ prefactor in the expression for $g_{\Delta,l}$. This
normalization is more convenient for the following reason. In the new
convention all conformal blocks are positive\ when operators are inserted at
the vertices of a square in the shown order (this corresponds to $z=\bar
{z}=1/2$). This is just as it should be, because this configuration is
reflection-positive in the Osterwalder-Shrader sense with respect to the
vertical median line (notice that the fields in the two sides of the
correlator are complex-conjugate of each other)\footnote{Actually, conformal
blocks are positive on the whole interval $0<z=\bar{z}<1$. Configurations
corresponding to such $z,\bar{z}$ can be mapped onto a rectangle, which is
reflection-positive.}. Thus $any$ s-channel contribution to the correlator,
even spin or odd, has to be positive. There is no disagreement with Doland and
Osborn \cite{do12}, because in their notation the extra minus sign would be
offset by a change in the sign of the OPE coefficient in the RHS of the correlator.

The (14)(23) channel $(\equiv$t-channel) conformal block decomposition can be
analyzed similarly. In this case we will need OPEs $\phi\times\phi^{\dagger}$
and $\chi\times\chi^{\dagger}$ and only fields appearing in both of these OPEs
will give a nonzero contribution, proportional to the product of the two OPE coefficients.

In \cite{r1} we have analyzed the particular case when $\phi$ is Hermitean and
$\chi=\phi$. In this case the s- and t-channels correspond to the same OPE
($\phi\times\phi$). In addition, only even spins contribute because of
permutation symmetry $x_{1}\leftrightarrow x_{2}$. Let us introduce the
notation for the sum of all s-channel contributions:%
\begin{equation}
G^{\text{+}}=\sum_{l\text{ even;}\Delta}p_{\Delta,l}g_{\Delta,l}(u,v)\,,
\label{eq:geven}%
\end{equation}
$(+$ means that we are summing over even spins only) and a tilde notation for
a contribution of the same set of operators in the t-channel:%
\begin{equation}
\widetilde{G}^{\text{+}}=G_{u\leftrightarrow v}^{+}=\sum p_{\Delta,l}%
g_{\Delta,l}(v,u)\,.
\end{equation}
Here we used the fact that going from the s- to the t-channel, which means
simply rotating the picture by $90^{%
%TCIMACRO{\U{b0}}%
%BeginExpansion
{{}^\circ}%
%EndExpansion
}$, interchanges $u$ and $v$. In this notation the crossing symmetry
constraint of \cite{r1} is written compactly as:%
\begin{equation}
G^{+}=\left(  \frac{u}{v}\right)  ^{d}\widetilde{G}^{+}\,.
\end{equation}
The appearance of the ($u/v)^{d}$ factor in this relation is due to a
nontrivial transformation of the prefactor $1/(x_{12}^{2d}x_{34}^{2d})$ in
(\ref{eq:confblocks}) under crossing.

Finally, an important technical remark. Unlike in \cite{r1}, to extract full
information from the 4-point function (\ref{eq:confblocks}), we will have to
consider not only the s- and t-channel OPEs, but also the u-channel ones
(13)(24). Conformal blocks for such `diagonal' OPEs are related to the
nearest-neighbor conformal blocks discussed above by analytic continuation,
which introduces spin-dependent signs into the crossing-symmetry constraints.
A useful way to keep track of these signs is not to consider the u-channel OPE
directly, but to instead apply the s- and t-channel decompositions to the
4-point function with the permuted insertion points:
\begin{equation}
\left\langle
\begin{array}
[c]{cc}%
\phi(x_{1})_{\bullet} & _{\bullet}\chi(x_{4})\\
\chi^{\dagger}(x_{2})^{\bullet} & ^{\bullet}\phi^{\dagger}(x_{3})
\end{array}
\right\rangle
\end{equation}
Here, we transposed the fields in the right side of the correlator. Now in the
t-channel we have the same OPE as we would have in the u-channel in
(\ref{eq:confblocks}). And in the s-channel we have the same OPE\ as in
(\ref{eq:confblocks}), except for the transposition. This transposition is
taken into account by reversing the sign of the odd-spin contributions in the
s-channel (and permuting the flavor indices accordingly, see below).

\subsection{Fundamental of $SO(N)$}

As a first example we will now consider the $SO(N)$ global symmetry case, with
a scalar primary operator $\phi_{a}$ transforming in the fundamental
representation. We normalize the 2-point function of $\phi_{a}$ as
$\left\langle \phi_{a}(x)\phi_{b}(0)\right\rangle =\delta_{ab}\left(
x^{2}\right)  ^{-d}$, $d=d_{\phi}$. Consider the 4-point function%
\begin{equation}
\left\langle
\begin{array}
[c]{cc}%
\phi_{a}\,_{\bullet} & _{\bullet}\phi_{d}\\
\phi_{b}\,^{\bullet} & ^{\bullet}\phi_{c}%
\end{array}
\right\rangle \equiv\frac{1}{x_{12}^{2d}x_{34}^{2d}}\,\mathcal{G}\Bigl[\!%
\begin{array}
[c]{c}%
a\ d\\[-5pt]%
b\ c
\end{array}
\Bigr|\,u,v\Bigr]\,.
\end{equation}
The operator insertion points are assumed numbered in the same order as in
(\ref{eq:confblocks}).

Operators appearing in the $\phi_{a}\times\phi_{b}$ OPE can transform under
the global symmetry as singlets $S$, symmetric traceless tensors $T_{(ab)}$,
or antisymmetric tensors $A_{[ab]}$:
\begin{align}
\phi_{a}\times\phi_{b}  &  =\delta_{ab}\mathds{1}\label{eq:son-ope}\\
&  +\delta_{ab}S^{(\alpha)}\quad\text{(even spins)}\nonumber\\
&  +T_{(ab)}^{(\alpha)}\quad\text{(even spins)}\nonumber\\
&  +A_{[ab]}^{(\alpha)}\quad\text{(odd spins)}\nonumber\,.
\end{align}
The index $\left(  \alpha\right)  $ shows that an arbitrary number of
operators of each type may in general be present, of various dimensions
$\Delta$ and spins $l$. However, permutation symmetry of the $\phi_{a}\phi
_{b}$ state implies that the spins of the $S$'s and $T$'s will be even, while
they will be odd for the $A$'s.

We note in passing that the stress tensor will be an $S$ of $\Delta=4,l=2$,
while the conserved $SO(N)$ current will be an $A$ of $\Delta=3$,$l=1$. The
OPE coefficients of these operators are related to the stress tensor and the
current central charges by the Ward identities \cite{op}, which should allow
to derive various bounds on these quantities by the method of \cite{cr}. The
simplest cases of these bounds have already been explored in \cite{r3}%
,\cite{poland}. In this paper we will not be making any assumptions about the
central charges of the theory and will treat the stress tensor and the current
on equal footing with all the other fields. (However, in future studies
central charge information may be useful; see the discussion Section.)

On the other hand, it will be important for us that the unit operator
$\mathds{1}$\ is always present in the $\phi_{a}\times\phi_{b}$ OPE, with a
unit coefficient.

As mentioned in the Introduction, we are interested to learn something about
the dimension of the lowest-dimension singlet scalar $($an $S$ of $l=0$). This
will require disentangling its contribution to the 4-point function from the
possibly present low-dimension scalars of type $T$.

We will now see what the crossing symmetry says about the relative
weights of various contributions in the $\phi\times\phi$ OPE.
Applying the conformal block decomposition in the s-channel we get:
\begin{equation}
\mathcal{G}\Bigl[\!%
\begin{array}
[c]{c}%
a\ d\\[-5pt]%
b\ c
\end{array}
\Bigr]\,=\raisebox{12pt}{$\xymatrix@=18pt{\ar@{{*}-{*}}[d]& \\ &
\ar@{{*}-{*}}[u] }$}\cdot(1+G_{S})+\left(
\raisebox{12pt}{$\xymatrix@=18pt{\ar@{{*}-{*}}[r]&
\\\ar@{{*}-{*}}[r]&
}$}+\raisebox{12pt}{$\xymatrix@=18pt{\ar@{{*}-{*}}[rd]&
\\\ar@{{*}-{*}}[ru]& }$}-\frac{2}{N}%
\raisebox{12pt}{$\xymatrix@=18pt{\ar@{{*}-{*}}[d]& \ar@{{*}-{*}}[d] \\& }$}\right)
\cdot G_{T}+\left(
\raisebox{12pt}{$\xymatrix@=18pt{\ar@{{*}-{*}}[r]& \\\ar@{{*}-{*}}[r]& }$}-\raisebox{12pt}{$\xymatrix@=18pt{\ar@{{*}-{*}}[rd]& \\\ar@{{*}-{*}}[ru]& }$}\right)
\cdot G_{A}\,. \label{eq:son-s}%
\end{equation}
Here $G_{S,T,A}$ are defined as in (\ref{eq:geven}), and sum up conformal
blocks of all fields of a given symmetry. Remember that $G_{S,T}$ contain only
even spins, while $G_{A}$ only the odd ones. The unit operator contributes
together with the singlets, and its conformal block is $\equiv1$. To keep
track of the index structure, we are using the graphical notation for tensors.
Every line means that the corresponding indices are contracted with the
$\delta$ tensor:%
\begin{equation}
\raisebox{12pt}{$\xymatrix@=18pt{\ar@{{*}-{*}}[d]& \\ & \ar@{{*}-{*}}[u] }$}=\delta
_{ab}\,\delta_{cd}\,,\text{ etc.}%
\end{equation}
The index structure of the symmetric traceless and the antisymmetric tensor
contributions in (\ref{eq:son-s}) is fixed by the symmetry (and by the
tracelessness, in the case of $G_{T}$). The signs are fixed from the
requirement that for $a=d\neq b=c$ all contributions have to be positive by
reflection positivity, see Section \ref{sec:conventions}. Apart from the sign
and the index structure, we do not keep track of the overall, positive,
normalization of each term. In other words, we know that each $G$ contains
conformal blocks summed with positive coefficients, but we do not keep track
of the normalization of these coefficients. This is sufficient for deriving
constraints on the operator spectrum, which is the focus of this paper. On the
other hand, normalization conventions will be important for any future studies
of the OPE coefficients.

Next we apply the t-channel conformal block decomposition to the same 4-point
function, and we get an alternative representation:%
\[
\mathcal{G}\Bigl[\!%
\begin{array}
[c]{c}%
a\ d\\[-5pt]%
b\ c
\end{array}
\Bigr]=\left(  \frac{u}{v}\right)  ^{d}\left\{
\raisebox{12pt}{$\xymatrix@=18pt{\ar@{{*}-{*}}[r]& \\\ar@{{*}-{*}}[r]& }$}\cdot
(1+\widetilde{G}_{S})+\left(
\raisebox{12pt}{$\xymatrix@=18pt{\ar@{{*}-{*}}[d]& \\ & \ar@{{*}-{*}}[u] }$}+\raisebox{12pt}{$\xymatrix@=18pt{\ar@{{*}-{*}}[rd]& \\\ar@{{*}-{*}}[ru]& }$}-\frac
{2}{N}%
\raisebox{12pt}{$\xymatrix@=18pt{\ar@{{*}-{*}}[r]&
\\\ar@{{*}-{*}}[r]& }$}\right) \cdot\widetilde{G}_{T}+\left(
\raisebox{12pt}{$\xymatrix@=18pt{\ar@{{*}-{*}}[d]& \\ &
\ar@{{*}-{*}}[u]
}$}-\raisebox{12pt}{$\xymatrix@=18pt{\ar@{{*}-{*}}[rd]&
\\\ar@{{*}-{*}}[ru]& }$}\right) \cdot\widetilde{G}_{A}\right\}\,.
\]
Note that to get this equation requires only changing the index structure
appropriately, permuting $u\leftrightarrow v$ (here we are using the tilde
notation introduced in Section \ref{sec:conventions}), and multiplying by
$\left(  u/v\right)  ^{d}$ to take into account how the $1/(x_{12}^{2d}%
x_{34}^{2d})$ transforms.

Now we equate the s- and t-channel representations and pick up coefficients
before each of the 3 inequivalent tensor structures:
$\raisebox{12pt}{$\xymatrix@=18pt{\ar@{{*}-{*}}[rd]& \\\ar@{{*}-{*}}[ru]& }$},\raisebox{12pt}{$\xymatrix@=18pt{\ar@{{*}-{*}}[d]& \\ & \ar@{{*}-{*}}[u] }$},\raisebox{12pt}{$\xymatrix@=18pt{\ar@{{*}-{*}}[r]& \\\ar@{{*}-{*}}[r]& }$}.$
We get 2 independent equations:%
\begin{subequations}
\begin{align}
u^{-d}\left\{  G_{T}-G_{A}\right\}   &  =v^{-d}\left\{  \widetilde{G}%
_{T}-\widetilde{G}_{A}\right\}  \,,\label{eq:son1}\\
u^{-d}\left\{  1+G_{S}-\frac{2}{N}G_{T}\right\}   &  =v^{-d}\left\{
\widetilde{G}_{T}+\widetilde{G}_{A}\right\}  \,, \label{eq:son2}%
\end{align}
and a third one which can be obtained from the second by $u\leftrightarrow v$:%
\end{subequations}
\begin{equation}
v^{-d}\left\{  1+\widetilde{G}_{S}-\frac{2}{N}\widetilde{G}_{T}\right\}
=u^{-d}\left\{  G_{T}+G_{A}\right\}  \,, \label{eq:son3}%
\end{equation}

Notice that for the $SO(N)$ case using the u-channel OPE would not yield any
new equation.

It will be convenient to rewrite the system (\ref{eq:son1}),(\ref{eq:son2}) in
the following equivalent form:%
\begin{subequations}
\begin{align}
F_{T}-F_{A}  &  =0\,,\label{eq:sonsr1}\\
F_{S}+\left(  1-\frac{2}{N}\right)  F_{T}+F_{A}  &  =1\,,\label{eq:sonsr2}\\
H_{S}-\left(  1+\frac{2}{N}\right)  H_{T}-H_{A}  &  =-1\,, \label{eq:sonsr3}%
\end{align}
where we introduced notation for (anti)symmetric linear combinations of $G$
and $\widetilde{G}$:%
\end{subequations}
\begin{align}
F(u,v)  &  =\frac{u^{-d}G(u,v)-v^{-d}G(v,u)}{v^{-d}-u^{-d}},\nonumber\\
H(u,v)  &  =\frac{u^{-d}G(u,v)+v^{-d}G(v,u)}{u^{-d}+v^{-d}}\,. \label{eq:FH}%
\end{align}
Thus (\ref{eq:sonsr1}) is obtained from (\ref{eq:son1}) just by grouping and
dividing by $v^{-d}-u^{-d}$. Eq.~(\ref{eq:sonsr2}) is obtained by taking the
difference of (\ref{eq:son2}) and (\ref{eq:son3}) and moving the contribution
of the unit operator to the RHS. Finally, Eq.~(\ref{eq:sonsr3}) follows by
taking the sum of (\ref{eq:son2}) and (\ref{eq:son3}), and again separating
the unity contribution.

Note that the functions $F(u,v)$ were already used in our previous work, while
the appearance of $H(u,v)$ is a new feature of the global symmetry analysis.
Writing the equations in terms of these functions is convenient because they
are highly symmetric with respect to the $z=\bar{z}=1/2$ point (they have only
even derivatives in $z+\bar{z}$ and $z-\bar{z}$ at this point).

The system (\ref{eq:sonsr1})-(\ref{eq:sonsr3}) is then the main
result of this Section. In an expanded notation, it can be written
as a `vectorial sum rule':%
\begin{equation}
\sum p_{\Delta,l}^{S}\left(
\begin{array}
[c]{c}%
0\\
F_{\Delta,l}\\
H_{\Delta,l}%
\end{array}
\right)  +\sum p_{\Delta,l}^{T}\left(
\begin{array}
[c]{c}%
F_{\Delta,l}\\
\left(  1-\frac{2}{N}\right)  F_{\Delta,l}\\
-\left(  1+\frac{2}{N}\right)  H_{\Delta,l}%
\end{array}
\right)  +\sum p_{\Delta,l}^{A}\left(
\begin{array}
[c]{c}%
-F_{\Delta,l}\\
F_{\Delta,l}\\
-H_{\Delta,l}%
\end{array}
\right)  =\left(
\begin{array}
[c]{c}%
0\\
1\\
-1
\end{array}
\right)\,.  \label{eq:son-vect}%
\end{equation}
Here the functions $F_{\Delta,l}(u,v)$ and $H_{\Delta,l}(u,v)$ are
related to the individual conformal blocks $g_{\Delta,l}$ by the
same formulas as $F$ and $H$ are related to $G$. Their dependence on
$d$ is left implicit. In each sum we are summing vector-functions
corresponding to the dimensions and spins present in this symmetry
channel, with positive coefficients. The total must converge to the
constant vector in the RHS.

Consequences of this new sum rule for the lowest singlet dimension will be
discussed below. Let us do however a quick counting of degrees of freedom. In
total we have three $G$-functions: $G_{S}$,$G_{T}$,$G_{A}$, each of which is
restricted only to the odd or even spins. The vectorial sum rule gives three
equations for their (anti)symmetric combinations $F$ and $H$. This coincidence
between the number of equations and unknowns is not accidental; see Section
\ref{sec:gen}. One may hope that the constraining power is similar to the case
without global symmetry, when we had one equation for only one function
$G^{+}$. We will see in Section \ref{sec:bounds} how this hope is realized.

\subsection{$U(1)$}

\label{sec:U(1)}

We next discuss the $U(1)$ global symmetry, as a case intermediate
between $SO(N)$ and $SU(N).$ On the one hand, we will be able to
check that the $U(1)$ constraints agree with the already considered
$SO(N)$ case for $N=2.$ On the other hand, the derivation will be
similar to the $SU(N)$ case which follows. In particular, we will be
working with complex fields and will need the u-channel OPE.

We want to derive constraints from crossing in the 4-point function of a
charge 1 complex scalar $\phi.$ Charge normalization is unimportant. The
nonvanishing correlators must have zero total charge, thus we are led to
consider $\left\langle \phi\phi\phi^{\dagger}\phi^{\dagger}\right\rangle .$
There are two basic OPEs:
\begin{align}
\text{Charge }0\text{ sector}  &  \text{:\quad}\phi\times\phi^{\dagger
}=\mathds{1}+\text{spins 0,1,2}\ldots,\label{eq:u1-ope0}\\
\text{Charge }2\text{ sector}  &  \text{:\quad}\phi\times\phi=\text{even spins
only\thinspace. } \label{eq:u1-ope2}%
\end{align}
Let us begin by considering the configuration%
\begin{equation}
\left\langle
\begin{array}
[c]{cc}%
\phi^{~}{}_{\bullet} & _{\bullet}\phi^{\dagger}\\
\phi^{\dagger\bullet} & ^{\bullet}\phi^{~}%
\end{array}
\right\rangle \,, \label{eq:u1-conf1}%
\end{equation}
which is the same as in (\ref{eq:confblocks}) for $\chi=\phi$. By doing the s-
and t-channel conformal block decompositions and demanding that the answers
agree we get a constraint:%
\begin{subequations}
\begin{equation}
u^{-d}\left\{  1+G_{0}^{+}+G_{0}^{-}\right\}  =v^{-d}\left\{  1+\widetilde
{G}_{0}^{+}+\widetilde{G}_{0}^{-}\right\}  \,. \label{eq:u(1)-1}%
\end{equation}
Here the subscript $0$ refers to the charge $0$ fields appearing in the
relevant $\phi\times\phi^{\dagger}$ OPE. As indicated in (\ref{eq:u1-ope0}),
this OPE contains both even and odd spin fields, whose contributions we
separate in $G_{0}^{\pm}$. According to the discussion in Section
\ref{sec:conventions}, reflection positivity of (\ref{eq:u1-conf1}) implies
that even and odd spins contribute in (\ref{eq:u(1)-1}) with the same positive sign.

Next consider the configuration with the transposed right side of the
correlator:%
\[
\left\langle
\begin{array}
[c]{cc}%
\phi^{~}{}_{\bullet} & _{\bullet}\phi^{~}\\
\phi^{\dagger\bullet} & ^{\bullet}\phi^{\dagger}%
\end{array}
\right\rangle \,.
\]
Equating the s- and t-channel decompositions we get:%
\begin{equation}
u^{-d}\left\{  1+G_{0}^{+}-G_{0}^{-}\right\}  =v^{-d}\widetilde{G}_{2}^{+}\,.
\label{eq:u(1)-2}%
\end{equation}
The LHS of this equation differs from the LHS of (\ref{eq:u(1)-1}) only by the
reversed sign of the odd spin contribution (see Section \ref{sec:conventions}%
). The t-channel decomposition appearing in the RHS is positive since the
configuration is reflection-positive in this direction.

Eqs.~(\ref{eq:u(1)-1}),(\ref{eq:u(1)-2}) solve the problem of expressing
crossing constraints in a $U(1)$ symmetric theory. Very recently, the same
equations also appeared in \cite{poland}. The authors of \cite{poland} have
noticed that they could get a bound on the lowest dimension singlet by using
just Eq.~(\ref{eq:u(1)-1}) (they only computed the bound in the supersymmetric
case, but the general case must be similar). Dropping the other equation
simplified the problem, but the downside was that they had to Taylor-expand up
to a pretty high order $(12)$ to extract the bound. Below we will show that if
one uses all equations the second-order expansion is already sufficient to
extract a bound.

Upon identification%
\end{subequations}
\begin{equation}
G_{S}=G_{0}^{+}\text{, \quad}G_{A}=G_{0}^{-},\quad G_{T}=\frac{1}{2}G_{2}^{+}%
\end{equation}
the $U(1)$ constraints become equivalent to the $N=2$ case of the $SO(N)$
constraints discussed above. The appearance of a positive factor $1/2$ is
consistent with the fact that we are keeping careful track of positivity but
not of the normalization.

\subsection{Fundamental of $SU(N)$}

Our last example is the $SU(N)$ case, with a primary scalar $\phi_{i}$
transforming in the fundamental. We have two basic OPEs:%
\begin{align}
\phi_{i}\times\phi_{\bar{\imath}}^{\dagger}  &  =\delta_{i\bar{\imath}%
}\mathds{1}+\delta_{i\bar{\imath}}\times\text{Singlets}(\text{spins
0,1,2\ldots})+\text{Adjoints}(\text{spins 0,1,2\ldots})\,, \label{eq:sun-ope1}%
\\
\phi_{i}\times\phi_{j}  &  =\yng(2)\text{~'s (even spins)}+\yng(1,1)\,\text{'s
(odd spins)}\,. \label{eq:sun-ope2}%
\end{align}
The representation content of the first OPE is $N\otimes\bar{N}=1+$Adj. Notice
that, in general, there will be singlets and adjoints of any spin. The adjoint
sector will contain the conserved current, but at present we are not using
information about its coefficient. The second OPE contains symmetric and
antisymmetric tensors, of even and odd spins respectively.

The constraints are now derived by a combination of what we did for $SO(N)$
and $U(1).$ First consider the following 4-point function configuration:%
\[
\left\langle
\begin{array}
[c]{cc}%
\phi_{i}{}_{\bullet} & _{\bullet}^{~}\phi_{\bar{\jmath}}^{\dagger}\\
\phi_{\bar{\imath}}^{\dagger\bullet} & ^{\bullet}\phi_{j}%
\end{array}
\right\rangle \,.
\]
The s- and t-channel conformal block decompositions are evaluated using the
first OPE. Equating them, we get a constraint:%
\begin{multline}
u^{-d}\left\{
\raisebox{12pt}{$\xymatrix@=18pt{\ar@{{*}-{o}}[d]& \\ & \ar@{{*}-{o}}[u] }$}(1+G^{+}%
_{S}+G^{-}_{S})+\left(  \raisebox{12pt}{$\xymatrix@=18pt{\ar@{{*}-{o}}[r]&
\\\ar@{{o}-{*}}[r]& }$}-\frac{1}{N}%
\raisebox{12pt}{$\xymatrix@=18pt{\ar@{{*}-{o}}[d]& \ar@{{o}-{*}}[d]
\\& }$}\right)  (G^{+}_{\text{Adj}}+G^{-}_{\text{Adj}})\right\} \nonumber\\
=v^{-d}\left\{  \raisebox{12pt}{$\xymatrix@=18pt{\ar@{{*}-{o}}[r]&
\\\ar@{{o}-{*}}[r]& }$}(1+\widetilde{G}^{+}_{S}+\widetilde{G}^{-}_{S})+\left(
\raisebox{12pt}{$\xymatrix@=18pt{\ar@{{*}-{o}}[d]& \ar@{{o}-{*}}[d]
\\& }$}-\frac{1}{N}\raisebox{12pt}{$\xymatrix@=18pt{\ar@{{*}-{o}}[r]&
\\\ar@{{o}-{*}}[r]& }$}\right)  (\widetilde{G}^{+}_{\text{Adj}}+\widetilde
{G}^{-}_{\text{Adj}})\right\}
\end{multline}
Here lines denote $SU(N)$-invariant contractions of $N$ (dots) and $\bar{N}$
(circles) indices by $\delta_{i\bar{\imath}}$. The tensor structure of the Adj
contributions is fixed by the tracelessness condition of the $SU(N)$
generators. The sign is fixed by the condition that for $i=\bar{\jmath}\neq
j=\bar{\imath}$ the s-channel contributions must be positive by reflection positivity.

Setting equal the coefficients before
$\raisebox{12pt}{$\xymatrix@=18pt{\ar@{{*}-{o}}[d]& \\ & \ar@{{*}-{o}}[u] }$}$
and
$\raisebox{12pt}{$\xymatrix@=18pt{\ar@{{*}-{o}}[r]& \\\ar@{{o}-{*}}[r]& }$}$
we get two equations:%
\begin{subequations}
\begin{equation}
u^{-d}\left\{  1+G^{+}_{S}+G^{-}_{S}-\frac{1}{N}(G^{+}_{\text{Adj}}%
+G^{-}_{\text{Adj}})\right\}  =v^{-d}\left\{  \widetilde{G}^{+}_{\text{Adj}%
}+\widetilde{G}^{-}_{\text{Adj}}\right\}  , \label{eq:SUN-1}%
\end{equation}
and a second one which is just the $u\leftrightarrow v$ version of the first.

Next we consider the transposed 4-point configuration:%
\[
\left\langle
\begin{array}
[c]{cc}%
\phi_{i}{}_{\bullet} & _{\bullet}^{~}\phi_{j}\\
\phi_{\bar{\imath}}^{\dagger\bullet} & ^{\bullet}\phi_{\bar{\jmath}}^{\dagger}%
\end{array}
\right\rangle \,.
\]
Equating the s- and t-channel decompositions, we get:%
\begin{multline*}
u^{-d}\left\{
\raisebox{12pt}{$\xymatrix@=18pt{\ar@{{*}-{o}}[d]& \\ & \ar@{{o}-{*}}[u] }$}(1+G_{S}%
^{+}-G_{S}^{-})+\left(  \raisebox{12pt}{$\xymatrix@=18pt{\ar@{{*}-{o}}[rd]&
\\\ar@{{o}-{*}}[ru]& }$}-\frac{1}{N}%
\raisebox{12pt}{$\xymatrix@=18pt{\ar@{{*}-{o}}[d]& \ar@{{*}-{o}}[d] \\& }$}\right)
(G_{\text{Adj}}^{+}-G_{\text{Adj}}^{-})\right\} \\
=v^{-d}\left\{  \left(  \raisebox{12pt}{$\xymatrix@=18pt{\ar@{{*}-{o}}[d]&
\ar@{{*}-{o}}[d] \\&
}$}+\raisebox{12pt}{$\xymatrix@=18pt{\ar@{{*}-{o}}[rd]&
\\\ar@{{o}-{*}}[ru]& }$}\right)  \widetilde{G}_{{\tiny \yng(2)}}+\left(
\raisebox{12pt}{$\xymatrix@=18pt{\ar@{{*}-{o}}[d]& \ar@{{*}-{o}}[d]
\\& }$}-\raisebox{12pt}{$\xymatrix@=18pt{\ar@{{*}-{o}}[rd]&
\\\ar@{{o}-{*}}[ru]& }$}\right)  \widetilde{G}_{{\tiny \yng(1,1)}}\right\}
\end{multline*}
The s-channel decomposition is obtained from the previous case by transposing
the index structure \textit{and} flipping the sign of the odd-spin
contributions. The t-channel decomposition is obtained by using the second OPE
(\ref{eq:sun-ope2}). The index structure is fixed by (anti)symmetry of the
exchanged fields, while the signs are determined by demanding positive
contributions for $i=\bar{\imath}\neq j=\bar{\jmath}$ (which makes the
configuration reflection-positive in the t-channel).

Collecting coefficients before the two inequivalent tensor structures, we get
two more equations, which this time are independent:%
\begin{align}
&  u^{-d}\left\{  1+G^{+}_{S}-G^{-}_{S}-\frac{1}{N}G^{+}_{\text{Adj}}+\frac
{1}{N}G^{-}_{\text{Adj}}\right\}  =v^{-d}\left\{  \widetilde{G}%
_{{\tiny \yng(2)}}+\widetilde{G}_{{\tiny \yng(1,1)}}\right\}
\,,\label{eq:SUN2}\\
&  u^{-d}\left\{  G^{+}_{\text{Adj}}-G^{-}_{\text{Adj}}\right\}
=v^{-d}\left\{  \widetilde{G}_{{\tiny \yng(2)}}-\widetilde{G}%
_{{\tiny \yng(1,1)}}\right\}  \,. \label{eq:SUN3}%
\end{align}

The system (\ref{eq:SUN-1})-(\ref{eq:SUN3}) solves the problem of expressing
the crossing symmetry constraints. Like in the $SO(N)$ case, we will find it
convenient to rewrite it by separating the unit operator contributions and
(anti)symmetrizing with respect to $u$ and $v$. We end up with the following
equivalent vectorial sum rule:%
\end{subequations}
\begin{equation}%
\begin{array}
[c]{rrrrrrcr}%
F_{S}^{+} & +F_{S}^{-} & +\left(  1-\frac{1}{N}\right)  F_{\text{Adj}}^{+} &
+\left(  1-\frac{1}{N}\right)  F_{\text{Adj}}^{-} &  &  & = & 1\\
H_{S}^{+} & +H_{S}^{-} & -\left(  1+\frac{1}{N}\right)  H_{\text{Adj}}^{+} &
-\left(  1+\frac{1}{N}\right)  H_{\text{Adj}}^{-} &  &  & = & -1\\
F_{S}^{+} & -F_{S}^{-} & -\frac{1}{N}F_{\text{Adj}}^{+} & +\frac{1}%
{N}F_{\text{Adj}}^{-} & +F_{{\tiny \yng(2)}} & +F_{{\tiny \yng(1,1)}} & = &
1\\
H_{S}^{+} & -H_{S}^{-} & -\frac{1}{N}H_{\text{Adj}}^{+} & +\frac{1}%
{N}H_{\text{Adj}}^{-} & -H_{{\tiny \yng(2)}} & -H_{{\tiny \yng(1,1)}} & = &
-1\\
&  & F_{\text{Adj}}^{+} & -F_{\text{Adj}}^{-} & +F_{{\tiny \yng(2)}} &
-F_{{\tiny \yng(1,1)}} & = & 0\\
&  & H_{\text{Adj}}^{+} & -H_{\text{Adj}}^{-} & -H_{{\tiny \yng(2)}} &
+H_{{\tiny \yng(1,1)}} & = & 0\,.
\end{array}
\label{eq:sun-rule}%
\end{equation}
Just like for $SO(N)$, the number of components, six, is equal to
the number of the OPE channels classified by
representation$\times$(spin parity): $S^{\pm}$, Adj$^{\pm}$, ${\tiny
\yng(2)\,}$, ${\tiny \yng(1,1)\,}$.

\subsection{General case}

\label{sec:gen}

In this Section we will consider the case of an arbitrary global symmetry
group $G$, with $\phi_{\alpha}$ transforming in an irreducible representation
$R.$ We aim at a general analysis of crossing symmetry constraints. In
particular, we would like to understand why the number of constraints came out
equal to the number of unknown functions in the explicit $SO(N)$ and $SU(N)$
examples above.

We will assume that $R$ is complex. The case of $R$ real is analogous but
simpler; necessary changes will be indicated below.

To understand the group theory aspect of the problem, we begin by
counting the number of scalar invariants which can be made out of
two $\phi$'s and two $\phi^{\dagger}$'s. These invariants can be
constructed by decomposing the products
$\phi_{\alpha}\times\phi_{\bar{\alpha}}^{\dagger}$ and $\phi_{\beta
}\times\phi_{\bar{\beta}}^{\dagger}$ into irreducible
representations and
contracting those. The tensor product representation decomposes as:%
\begin{equation}
R\otimes\bar{R}=%
{\displaystyle\bigoplus_{i=1}^{n}} {}\ r_{i}(+\bar{r}_{i})\,,
\end{equation}
where $(+\bar{r}_{i})$ indicates that the representations in the RHS must be
either real or come in complex conjugate pairs. To simplify the discussion,
assume for now that all $r_{i}$ are real and different. In accord with the
above decomposition, we have
\begin{equation}
\phi_{\alpha}\times\phi_{\bar{\alpha}}^{\dagger}=\sum_{i}\sum_{A_{i}}%
C_{\alpha\bar{\alpha}A_{i}}^{i}\Psi_{A_{i}}^{i}\,,
\end{equation}
where the objects $\Psi_{A_{i}}$ transform in the $r_{i}$, and $C_{\alpha
\bar{\alpha}A_{i}}^{i}$ are the Clebsch-Gordan coefficients ($A_{i}$ is the
index in the $r_{i}$). Then we can construct exactly $n$ invariant tensors by
contracting two Clebsch-Gordan coefficients:
\begin{equation}
T_{\alpha\bar{\alpha}\beta\bar{\beta}}^{i}=\sum_{A_{i}}C_{\alpha\bar{\alpha
}A_{i}}^{i}C_{\beta\bar{\beta}A_{i}}^{i}\,, \label{eq:CG}%
\end{equation}
so that the product of two $\phi$'s and two $\phi^{\dagger}$'s can be
decomposed into a sum of $T$'s:
\begin{align}
\phi_{\alpha}\phi_{\bar{\alpha}}^{\dagger}\phi_{\beta}\phi_{\bar{\beta}%
}^{\dagger}  &  =\sum_{i}\xi_{i}T_{\alpha\bar{\alpha}\beta\bar{\beta}}%
^{i}\nonumber\\
&  =\sum_{i}\tilde{\xi}_{i}T_{\alpha\bar{\beta}\beta\bar{\alpha}}^{i}\,,
\end{align}
where in the second line we indicated that we can do the same construction in
a crossed fashion, by starting with the $\phi_{\alpha}\times\phi_{\bar{\beta}%
}^{\dagger}$ product. The fact that both decompositions exist means that the
invariant tensors satisfy a linear relation (`Fierz identity')
\begin{equation}
T_{\alpha\bar{\alpha}\beta\bar{\beta}}^{i}=\mathcal{F}_{\,\,i^{\prime}}%
^{i}T_{\alpha\bar{\beta}\beta\bar{\alpha}}^{i^{\prime}}\,. \label{eq:F1}%
\end{equation}
The matrix $\mathcal{F}$ is invertible and must satisfy $\mathcal{F}%
^{2}=\mathds{1},$ since crossing is a $\mathds{Z}_{2}$ operation.

It is also possible to construct invariants by starting from $\phi_{\alpha
}\times\phi_{\beta}$, which requires the tensor product%
\begin{equation}
R\otimes R=%
%TCIMACRO{\dbigoplus _{j=1}^{n}}%
%BeginExpansion
{\displaystyle\bigoplus_{j=1}^{n}}
%EndExpansion
{}\ \tilde{r}_{j}.
\end{equation}
Assume for now that all $\tilde{r}$'s appearing in this product are also
distinct (excluding as well the possibility for the same representation to
occur both in the symmetric and antisymmetric part of the tensor product$)$.
Under this simplifying assumption, the number of $\tilde{r}$'s is the same as
the number of $r$'s. Indeed, we can construct invariant tensors%
\begin{equation}
\widetilde{T}_{\alpha\beta\bar{\alpha}\bar{\beta}}^{j}=\sum_{A_{j}}%
C_{\alpha\beta A_{j}}^{j}C_{\bar{\alpha}\bar{\beta}A_{j}}^{j}\,,
\label{eq:Ttilde}%
\end{equation}
where $C_{\alpha\beta A_{j}}^{j}$ (resp. $C_{\bar{\alpha}\bar{\beta}A_{j}}%
^{j}$) are the Clebsch-Gordan coefficients for $\tilde{r}_{j}$ in $R\times R$
(resp. $\overline{\tilde{r}}_{j}$ in $\bar{R}\times\bar{R}$). These must be
related to $T$'s by another Fierz identity%
\begin{equation}
T_{\alpha\bar{\alpha}\beta\bar{\beta}}^{i}=\widetilde{\mathcal{F}}_{\,\,j}%
^{i}\widetilde{T}_{\alpha\beta\bar{\alpha}\bar{\beta}}^{j}\,, \label{eq:F2}%
\end{equation}
where $\widetilde{\mathcal{F}}$ is again an invertible matrix. Notice however
that $T\neq\widetilde{T}$ and thus $\widetilde{\mathcal{F}}^{2}\neq
\mathds{1}.$

After this prelude, we come back to our problem of analyzing the crossing
symmetry constraints of the CFT 4-point function.

\textit{Step 1}. Let us compare the s- and t-channel conformal block
decompositions:%
\begin{equation}
\left\langle
\begin{array}
[c]{cc}%
\phi_{\alpha}{}_{\bullet} & _{\bullet}^{~}\phi_{\bar{\beta}}^{\dagger}\\
\phi_{\bar{\alpha}}^{\dagger\bullet} & ^{\bullet}\phi_{\beta}%
\end{array}
\right\rangle \,=\sum_{i}%
\raisebox{1.7em}{$\xymatrix@=10pt{{\alpha}\ar@{-}[rd]& &
&\bar{\beta}\ar@{-}[ld]
\\  & *+[o][F]{} \ar@{=}[r]^{\displaystyle r_i} &*+[o][F]{} &  \\
\bar{\alpha}\ar@{-}[ru]& & &\beta\ar@{-}[lu]}$}\,=\sum_{i}%
\raisebox{2.5em}{$\xymatrix@=10pt{\alpha\ar@{-}[rd]& &
\bar{\beta}\ar@{-}[ld]
\\  & *+[o][F]{} \ar@{=}[d]^{\displaystyle r_i}  &  \\
& *+[o][F]{} & \\
\bar{\alpha}\ar@{-}[ru]&  &\beta\ar@{-}[lu]}$}\,. \label{eq:conf1}%
\end{equation}
Introduce functions $G_{i}$ which sum up conformal blocks of operators in the
representation $r_{i}$ (which will in general occur in both even and odd
spins). The tensor structure of these contributions will be given precisely by
the invariant tensors $T$ introduced above. The crossing symmetry constraint
then takes the form:%
\begin{equation}
\sum_{i}T_{\alpha\bar{\alpha}\beta\bar{\beta}}^{i}G_{i}(u,v)=\sum_{i}%
T_{\alpha\bar{\beta}\beta\bar{\alpha}}^{i}G_{i}(v,u)\,. \label{eq:st-gen}%
\end{equation}
Here we assume that the signs of $T$'s have been chosen in agreement with
reflection positivity. To simplify the notation we included the $u^{-d}$,
$v^{-d}$ prefactors in the definition of $G_{i}$. We also do not separate the
unit operator explicitly.

Eq.~(\ref{eq:st-gen}) will be consistent with the first Fierz identity
(\ref{eq:F1}) if and only if%
\begin{equation}
G_{i}(u,v)=\mathcal{F}_{i}^{\,i^{\prime}}G_{i^{\prime}}(v,u)\,.
\end{equation}
Let us now define even and odd combinations:%
\begin{equation}
^{(\pm)}G_{i}=G_{i}(u,v)\pm G_{i}(v,u)\,. \label{eq:evenodd}%
\end{equation}
These are the analogues of the $F$ and $H$ functions from Eq.~(\ref{eq:FH}).
We put the index $\left(  \pm\right)  $ on the left to stress that it has
nothing to do with the spin parity index used in the previous Sections; these
functions receive contributions from both even and odd spins. We have%
\begin{equation}
(\mathds{P}_{\pm})_{i}^{\,\,i^{\prime}}G_{i^{\prime}}^{\left(  \pm\right)
}=0\,, \label{eq:P}%
\end{equation}
where $\mathds{P}_{\pm}=(1\mp\mathcal{F})/2$ are projectors, $\left(
\mathds{P}_{\pm}\right)  ^{2}=\mathds{P}_{\pm}$ by using $\mathcal{F}%
^{2}=\mathds{1}.$ Going to the diagonal basis for
$\mathds{P}_{\pm}$, it is clear that Eq.~(\ref{eq:P}) represents a
total of $n$ constraints.

\textit{Step 2}. We next compare the s- and t-channel conformal block
decompositions of the transposed 4-point function:%
\begin{equation}
\left\langle
\begin{array}
[c]{cc}%
\phi_{\alpha}{}_{\bullet} & _{\bullet}^{~}\phi_{\beta}\\
\phi_{\bar{\alpha}}^{\dagger\bullet} & ^{\bullet}\phi_{\bar{\beta}}^{\dagger}%
\end{array}
\right\rangle \,=\sum_{i}%
\raisebox{1.7em}{$\xymatrix@=10pt{{\alpha}\ar@{-}[rd]& &
&{\beta}\ar@{-}[ld]
\\  & *+[o][F]{} \ar@{=}[r]^{\displaystyle r_i} &*+[o][F]{} &  \\
\bar{\alpha}\ar@{-}[ru]& & &\bar{\beta}\ar@{-}[lu]}$}\,=\sum_{j}%
\raisebox{2.5em}{$\xymatrix@=10pt{\alpha\ar@{-}[rd]& &
{\beta}\ar@{-}[ld]
\\  & *+[o][F]{} \ar@{=}[d]^{\displaystyle \tilde{r}_j}  &  \\
& *+[o][F]{} & \\
\bar{\alpha}\ar@{-}[ru]&  &\bar{\beta}\ar@{-}[lu]}$}\,. \label{eq:conf2}%
\end{equation}
The crossing symmetry constraint can be written in terms of the invariant
tensors introduced above as:%
\begin{equation}
\sum_{i}T_{\alpha\bar{\alpha}\beta\bar{\beta}}^{i}[G_{i}^{+}(u,v)-G_{i}%
^{-}(u,v)]=\sum_{j}\widetilde{T}_{\alpha\beta\bar{\alpha}\bar{\beta}}^{j}%
G_{j}(v,u)\,. \label{eq:u-gen}%
\end{equation}
Here we have shown explicitly that the odd spin parts $G_{i}^{-}$ of the
$G_{i}$ flip signs compared to the above configuration (\ref{eq:conf1}). Note
as well that each of the functions $G_{j}$ will include even or odd spins
only, depending if $\tilde{r}_{j}$ occurs in the symmetric or antisymmetric
part of $R\times R$.

For Eq. (\ref{eq:u-gen}) to be consistent with the second Fierz identity
(\ref{eq:F2}), we must have%
\begin{equation}
G_{i}^{+}(u,v)-G_{i}^{-}(u,v)=\widetilde{\mathcal{F}}_{i}^{\,\,j}G_{j}(v,u)\,.
\end{equation}
Since the functions in the RHS and LHS now refer to completely different OPE
channels ($r_{i}$ in $\phi\times\phi^{\dagger}$ vs $\tilde{r}_{j}$ in
$\phi\times\phi$), this equation gives exactly $2n$ constraints when
(anti)symmetrizing in $u,v.$

To summarize, we expect $3n$ constraints for $3n$ channels $r_{i}^{\pm}$
,$\tilde{r}_{j}$. In particular, $n=3$ for the fundamental of $SU(N).$

In case when $R$ is a real representation, we only have one set of invariant
tensors, whose Fierz dictionary matrix satisfies $\mathcal{F}^{2}=\mathds{1}$.
In this case each of $n$ representations in the $R\times R$ product will
contribute with only even or odd spins. Only the first step of the above
analysis is needed in this case. We will get $n$ constraints for $n$ channels.
The fundamental of $SO(N)$ corresponds to $n=3.$

\textit{Generalizations.} Let us now discuss how one can relax the assumptions
on the content of $R\otimes\bar{R}$ and $R\otimes R$ taken in the above
argument. In general, $R\otimes\bar{R}$ may contain repetitions of the same
representation as well as conjugate pairs, while $R\otimes R$ may contain the
same representation in both symmetric $(s)$ and antisymmetric $(a)$ part. As
it will become clear below, these two things must happen simultaneously. A
sufficiently representative example is $R=\overline{15}$ of $G=SU(3)$
\cite{slansky}:%
\begin{align}
15\otimes\overline{15}  &  =1+64+(8_{1}+8_{2})+(27_{1}+27_{2})+(10+\overline
{10})+(35+\overline{35})\,,\\
\overline{15}\otimes\overline{15}  &  =3_{a}+\bar{6}_{s}+15_{s}^{\prime
}+24_{a}+42_{a}+60_{s}+(15_{s}+15_{a})+(21_{a}+24_{s})\,.
\end{align}
In $15\otimes\overline{15}$ we have $8$ and $27$ appearing twice each, and
also two conjugate pairs ($10+\overline{10}$ and $35+\overline{35}),$ while in
$\overline{15}\otimes\overline{15}$, $15$ and $24$ appear both as $s$ and $a.$

In cases like this, it is slightly more involved to count the quartic
invariants. When counting in the $R\otimes\bar{R}$ channel, every conjugate
pair $r+\bar{r}$ gives two invariants which for future purposes we
(anti)symmetrize with respect to $(\alpha\bar{\alpha})\leftrightarrow
(\beta\bar{\beta})$:%
\begin{equation}
\sum_{A}C_{\alpha\bar{\alpha}A}^{r}C_{\beta\bar{\beta}A}^{\bar{r}}\pm
C_{\alpha\bar{\alpha}A}^{\bar{r}}C_{\beta\bar{\beta}A}^{r}\,.
\label{eq:pairinvs}%
\end{equation}
In the same channel, a $k$-fold repetition of a real representation $r$ gives
rise to $k^{2}$ invariants:%
\begin{equation}
\sum_{A}C_{\alpha\bar{\alpha}A}^{r_{i}}C_{\beta\bar{\beta}A}^{r_{j}}%
\qquad(i,j=1\ldots k)\,,
\end{equation}
which can be (anti)symmetrized with respect to $(\alpha\bar{\alpha
})\leftrightarrow(\beta\bar{\beta})$, producing $k(k+1)/2$ symmetrics and
$k(k-1)/2$ antisymmetrics.

When counting in the $R\otimes R$ channel, every representation $r$
occurring
both as $s$ and $a$ gives rise to $4$ invariants%
\begin{equation}
\sum_{A}C_{\alpha\beta A}^{r_{s/a}}C_{\bar{\alpha}\bar{\beta}A}^{\bar{r}%
_{s/a}}\,,
\end{equation}
out of which two are symmetric and two antisymmetric in $(\alpha\bar{\alpha
})\leftrightarrow(\beta\bar{\beta})$.

The total number of invariants must of course be the same counted in
$R\otimes\bar{R}$ and in $R\otimes R$ channel. This is indeed true in the
above example, when both $15\otimes\overline{15}$ and $\overline{15}%
\otimes\overline{15}$ give $14$. The number of symmetric in
$(\alpha\bar{\alpha})\leftrightarrow(\beta\bar{\beta})$ invariants
also agrees $(10$ in both channels). This is also true in general.
An intuitive argument is as follows. The total number of invariants
equals the number of independent
coupling constants in the scalar potential $V(\phi_{1},\phi_{2}^{\dagger}%
,\phi_{3},\phi_{4}^{\dagger}$) where $\phi_{i}$ are four non-identical scalars
transforming in $R.$ This number should be the same whether you begin by
contracting $\phi_{1}$ with $\phi_{2}^{\dagger}$ or $\phi_{3}$. Analogously,
the number of \textit{symmetric} invariants is the number of quartic couplings
if we identify $\phi_{3}\equiv\phi_{1}$, $\phi_{4}\equiv\phi_{2}$.

Each of the two Fierz identities (\ref{eq:F1}) and (\ref{eq:F2}) will now
split into two, one for symmetric and one for antisymmetric invariants.

\begin{figure}
[h]
\begin{center}
\includegraphics[
height=1.6578in,
width=2.8632in
]%
{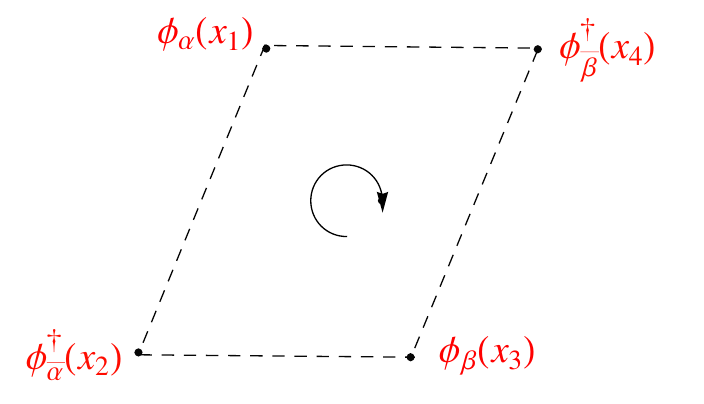}%
\caption{\textit{For any 4-point configuration, there exists a conformal
transformation which maps it onto a parallelogram.}}%
\label{fig:para}%
\end{center}
\end{figure}

Let us now proceed to the crossing symmetry analysis of the 4-point function
$\left\langle \phi_{\alpha}\phi_{\bar{\alpha}}^{\dagger}\phi_{\beta}\phi
_{\bar{\beta}}^{\dagger}\right\rangle .$ To begin with, out of all the
invariant tensors discussed above, only the symmetric ones will appear as the
coefficients in the conformal block expansions of this correlator\footnote{In
a general Lorentz-invariant theory, the flavor structure of this correlator
will involve both symmetric and antisymmetric tensors.}. The $(\alpha
\bar{\alpha})\leftrightarrow(\beta\bar{\beta})$ symmetry is made manifest by
applying a conformal transformation which maps a generic 4-point configuration
in (\ref{eq:conf1}) onto a parallelogram. The $180^{%
%TCIMACRO{\U{b0}}%
%BeginExpansion
{{}^\circ}%
%EndExpansion
}$ rotation symmetry of the parallelogram then acts on the indices
as $(\alpha\bar{\alpha})\leftrightarrow(\beta\bar{\beta}),$ see
Fig.~\ref{fig:para}.

To see how this symmetry arises in the conformal block decomposition, consider the OPE%
\begin{align}
\phi_{\alpha}\times\phi_{\bar{\alpha}}^{\dagger}  &  =\sum_{r\text{ real}%
}\left(  \sum_{i=1}^{k_{r}}\lambda_{O}^{i}C_{\alpha\bar{\alpha}A}^{r_{i}%
}\right)  O_{A}\\
&  +\sum_{r+\bar{r}\text{ pairs}}\lambda_{O}C_{\alpha\bar{\alpha}A}^{r}%
O_{A}+(-1)^{l}\lambda_{O}^{\ast}C_{\alpha\bar{\alpha}A}^{\bar{r}}%
O_{A}^{\dagger}\,.
\end{align}
Here in the first line we include all operators belonging to the real
representations. If the representation is repeated $k$ times in the
$R\otimes\bar{R}$ product, there are $k$ independent Clebsch-Gordan
coefficients, and $k$ independent real OPE coefficients $\lambda_{O}^{i}$. In
the second line we have operators from the complex-conjugate pairs, whose OPE
coefficients are always complex-conjugate, up to a spin-dependent minus sign.

By using this OPE in the s-channel conformal block decomposition of
(\ref{eq:conf1}), we see that indeed only symmetric invariant
tensors arise. Notice that the off-diagonal invariant tensors
($i\neq j$) in the case of
repeated representations will appear with coefficients$\ \lambda_{O}%
^{i}\lambda_{O}^{j}$ ($\times$conformal block), which are not positive
definite. We will discuss below what this means for the subsequent application
of the derived constraints.

We then consider the t-channel conformal block decomposition of
(\ref{eq:conf1}), and repeat the analysis of Step 1. The resulting
number of constraints is equal to the number $n_{\text{sym}}$ of
symmetric invariants,
while the number of representation$\times$(spin parity) channels is $2n_{\text{sym}%
}$.

To generalize Step 2, we have to consider the t-channel
decomposition of (\ref{eq:conf2}). In this channel, the OPE parity
selection rules imply immediately that only symmetric tensor
structures appear, in agreement with the above general result. If
the $s$ and $a$ representations are not repeated, as in the
$\overline{15}\otimes\overline{15}$ example, then only diagonal
terms are present, and all conformal blocks enter with positive
coefficients. Compared to Step 1, we will have $n_{\text{sym}}$ new
representation$\times$(spin parity) channels and $2n_{\text{sym
}}$new constraints.

We are done: we have a total $3n_{\text{sym}}$ constraints for $3n_{\text{sym}%
}$ channels. Moreover, these constraints distinguish not only
different representations appearing in the OPE, but also different
copies of the same representation, and how they `interfere' among
each other.

Let us now come back to the fact that if repeated representations
are present in $R\otimes\bar{R}$, the off-diagonal `interference'
channels have coefficients $\lambda_{O}^{i}\lambda_{O}^{j}$. To
appreciate the difficulty that this creates, readers unfamiliar with
our method of linear functionals are encouraged to read the rest of
this Section after having read Section \ref{sec:bounds}.

Consider then our abstract way (\ref{eq:vecspace}) of representing
the vectorial sum rule. It is crucial for us that when all
coefficients $p_{\alpha}$ are allowed to vary subject to the
positivity constraints $p_{\alpha}\geq0$, linear combinations in the
LHS fill a convex cone. In particular, this allows us to use the
dual formulation of the problem in the form (\ref{eq:lambda3}).
Since the off-diagonal coefficients may be negative, the geometric
interpretation in this case is not as obvious. Notice however that
the off-diagonal coefficients cannot become arbitrarily negative
since they are not independent of the diagonal ones. For a sharp
formulation, consider a
symmetric real matrix%
\begin{equation}
P_{ij}=\sum_{O}\lambda_{O}^{i}\lambda_{O}^{j}\,\text{,} \label{eq:pdrep}%
\end{equation}
where we allow for presence of more than one operator $O$ with a given
dimension, spin, and representation. The characterizing property of $P$ is
positive-definiteness:%
\begin{equation}
P_{ij}s_{i}s_{j}\geq0\qquad\forall s_{i}\in\mathds{R}\,. \label{eq:posdef}%
\end{equation}
Now, as can be seen from this equation, the set of positive-definite
matrices forms by itself a convex cone. It follows that the set of
vectors in the LHS of the vectorial sum rule will remain a convex
cone even if repeated representations are present. Constraints
(\ref{eq:posdef}) replace the simple inequality $p_{\alpha}\geq0.$
In practical applications these constraints may have to be
discretized by choosing a finite set of vectors $s_{i}$.

The dual formulation (\ref{eq:lambda3}) is extended to the present case as
follows. For the vectors $\mathbf{x}_{ij}$ in the LHS of the sum rule
corresponding to diagonal $(i=j)$ and off-diagonal ($i\neq j$) channels of the
repeated representation, the simple condition $\Lambda\lbrack\mathbf{x}%
_{\alpha}]\geq0$ must be replaced by the following condition on the matrix
$\Lambda\lbrack\mathbf{x}_{ij}]$:%
\begin{equation}
P_{ij}\Lambda\lbrack\mathbf{x}_{ij}]\geq0\quad\forall P\text{
positive-definite.}%
\end{equation}
In other words, $\Lambda\lbrack\mathbf{x}_{ij}]$ must belong to the cone
\textit{dual} to the cone of positive-definite matrices. However, the latter
cone is in fact self-dual, as can be easily inferred from the representation
(\ref{eq:pdrep}). Thus, $\Lambda\lbrack\mathbf{x}_{ij}]$ must be itself positive-definite.

In the above discussion, only real representations were allowed to repeat in
$R\otimes\bar{R}$. However, repetitions of complex pairs could be treated
similarly; the only difference is that the corresponding $P$ matrices will be
positive-definite Hermitean rather than real.

\section{Bounds on the lowest singlet scalar dimension}

\label{sec:bounds}

\subsection{Generalities}

\label{sec:generalities}

The previous Section would be a futile exercise in group theory if our
vectorial sum rules did not have any useful consequences. We will now discuss
how they can be used to bound the gap in the singlet scalar sector. Consider
the $SO(N)$ case for definiteness. Given a CFT spectrum, the sum rule
(\ref{eq:son-vect}) can be viewed as an equation for the coefficients
$p_{\Delta,l}^{S,T,A}$. If we start imposing restrictions on the spectrum,
such as raising the singlet scalar gap, it is conceivable that this equation
will not have any solution consistent with the positivity requirement
$p_{\Delta,l}\geq0$. This is in fact precisely what will happen.

An equivalent, geometric, way to view this is as follows. Let us rewrite the
sum rule (\ref{eq:son-vect}) schematically as an equation in a linear space
$V$ of functions from two variables $u,v$ into $\mathds{R}^{3}$ (vector space
of vector-functions):%
\begin{equation}
\sum p_{\alpha}\mathbf{x}_{\alpha}=\mathbf{y}\,. \label{eq:vecspace}%
\end{equation}
Here vectors $\mathbf{x}_{\alpha}$ represent all vector-functions appearing in
the LHS of (\ref{eq:son-vect}), while the $\mathbf{y}$ is the vector
corresponding to the RHS.

\begin{figure}[h]
\begin{center}
\includegraphics[height=1.5487in, width=1.7323in]{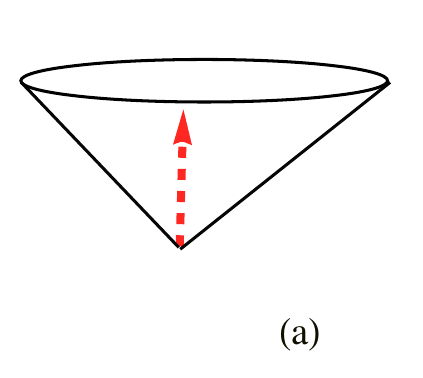}\hspace{1cm}
\includegraphics[]{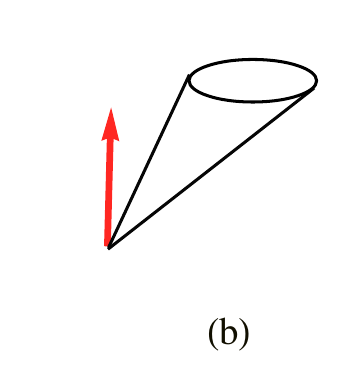}\hspace{1cm} \includegraphics[]{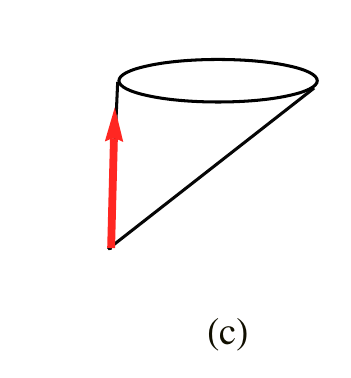}
\end{center}
\caption{\textit{Geometric interpretation of the sum rule: (a) the sum rule
has a solution $\Leftrightarrow$ $\mathbf{y}$ belongs to the cone; (b) the
assumed spectrum is such that the sum rule does not allow for a solution
$\Leftrightarrow$ $\mathbf{y}$ does not belong to the cone; (c) for }%
$\Delta_{S}=\Delta_{S}^{\text{cr}}$\textit{, the }$\mathbf{y}$\textit{ belongs
to the cone boundary.}}%
\label{fig:cone}%
\end{figure}

For a fixed CFT spectrum and varying $p_{\alpha}\geq0$, the vectors in the LHS
of (\ref{eq:vecspace}) fill a convex cone, so the question is whether the
vector $\mathbf{y}$ belongs to this cone. Imposing restrictions on the
spectrum reduces the set of vectors $\mathbf{x}_{\alpha}$ generating the cone,
and the cone shrinks. It may be that the new smaller cone no longer contains
$\mathbf{y}$, see Fig.~\ref{fig:cone}(a,b).

In this paper we will be dealing with two types of restrictions on the
spectrum. First of all, we will always impose the unitarity bounds
\cite{mack}
\begin{equation}
\Delta\geq1\quad(l=0)\,,\qquad\Delta\geq l+2\quad(l\geq1\text{)\thinspace.}%
\end{equation}
These lower bounds on operator dimension are a completely general property of
unitary 4D CFTs. Note that they depend only on spin and not, say, on the
global symmetry representation in which the operator transforms.

Second, we will impose a lower bound on the dimension of scalar singlets:%
\begin{equation}
\Delta\geq\Delta_{S}\quad(l=0\text{ singlets only).}%
\end{equation}
According to the above discussion, increasing $\Delta_{S}$ makes the cone
shrink. Our goal will be to show that for $\Delta_{S}$ above a certain
critical value $\Delta_{S}^{\text{cr}}$, the $\mathbf{y}$ is not in the cone.
This critical value will then be a theoretical upper bound on the dimension of
the first singlet scalar, valid in an arbitrary unitary 4D CFT. This is the
bound denoted by $f_{S}$ in Eq.~(\ref{eq:boundS}) of the introduction.

Note that the list of vectors generating the cone, and thus the cone itself,
vary continuously with $\Delta_{S}$. Since for $\Delta_{S}<\Delta
_{S}^{\text{cr}}$ the $\mathbf{y}$ is inside the cone and for $\Delta
_{S}>\Delta_{S}^{\text{cr}}$ it's outside, for $\Delta_{S}=\Delta
_{S}^{\text{cr}}$ it must belong to the cone boundary, see Fig.~\ref{fig:cone}(c).

Up to now we were keeping the external scalar dimension $d_{\phi}$ fixed.
However, the vectors entering the sum rule depend on $d_{\phi}$ via
Eq.~(\ref{eq:FH}). This dependence is continuous, and thus the cone will vary
continuously with $d_{\phi}$. It follows that the bound $f_{S}(d_{\phi})$, if
it exists, will have a continuous dependence on $d_{\phi}$.

The reader may be worried that the above discussion was not totally rigorous.
Indeed, the vector space $V$ is infinite-dimensional, and there may be
subtleties of convergence. However, below we will always be considering a
finite-dimensional subspace of $V$, by Taylor-expanding the sum rule up to a
fixed finite order $k$. On the one hand, this means that the bounds that we
will derive will not be optimal (they will approach optimality in the
$k\rightarrow\infty$ limit). On the other hand, finite-dimensional analysis is
simpler both practically and from the point of view of mathematical rigor. In
particular, the statement that $f_{S}$ is a continuous function of $d_{\phi}$
is safe at finite $k$.

\subsection{Best possible bound for $d_{\phi}=1$}

We will begin by analyzing carefully the case $d_{\phi}=1$. The reader may
wonder why this is necessary, since there is a theorem that a scalar
saturating the unitarity bound is necessarily free. This theorem is easy to
prove: starting from the 2-point function $\left\langle \phi(x)\phi
(0)\right\rangle =x^{-2}$ we deduce $\left\langle \partial^{2}\phi
(x)\partial^{2}\phi(0)\right\rangle =0$ and thus $\partial^{2}\phi=0$ in the
Hilbert space. In the $SO(N)$ case, we will have $N$ free real scalars
$\phi_{a}$. The $\phi_{a}\times\phi_{b}$ OPE will contain two scalar
operators
\begin{equation}
S^{(0)}=\text{:}\phi_{c}\phi_{c}\text{:,\quad}T_{(ab)}^{(0)}=\text{:}\phi
_{a}\phi_{b}-\frac{1}{N}\delta_{ab}(\phi_{c}\phi_{c})\text{:,}%
\end{equation}
the first a singlet, the second a symmetric traceless, both dimension $2$.
Bounding the singlet scalar dimension from above seems a waste of time: the
exact solution shows that it must be exactly $2$.

Yet it is not useless to give a derivation of such a result---that at
$d_{\phi}=1$ there must be a singlet scalar of dimension $\leq2$---using our
method based on the vectorial sum rule. The reason is very simple: unlike the
derivation sketched above, our method is robust with respect to small
variations in $d_{\phi}$. If we show that it works at $d_{\phi}=1$, it is
guaranteed to work as well for $d_{\phi}$ sufficiently close to $1$. Thus we
will know that a bound \textit{exists} for $d_{\phi}$ close to $1$ and, since
it is a continuous function of $d_{\phi}$, that it approaches 2 as $d_{\phi
}\rightarrow1$. We believe that such an existence proof is conceptually
important. It is also easier than actually computing the bound at $d_{\phi}%
>1$. The latter problem will be discussed in the next Section.

Consider then the sum rule (\ref{eq:son-vect}) for $d_{\phi}=1$. This sum rule
has at least one solution--the one corresponding to the theory of $N$ free
scalars. This solution is very special, in that among all the vectors
appearing in the RHS, only those corresponding to twist $\Delta-l=2$ fields
will have nonzero coefficients. This is because in free theory no fields of
other twists appear in the $\phi_{a}\times\phi_{b}$ OPE. Apart from the two
$\Delta=2$ scalars mentioned above, there are infinitely many twist $2$ fields
of the form%
\begin{equation}
\text{:}\phi_{a}\overleftrightarrow{\partial}_{\mu_{1}}\ldots
\overleftrightarrow{\partial}_{\mu_{l}}\phi_{b}\text{:\thinspace,}%
\end{equation}
appropriately (anti)symmetrized in $a$,$b$ to separate $SO(N)$
representations. Expanding the free field theory 4-point function%
\begin{equation}
\left\langle \phi_{a}(x_{1})\phi_{b}(x_{2})\phi_{c}(x_{3})\phi_{d}%
(x_{4})\right\rangle =\frac{\delta_{ab}\delta_{cd}}{x_{12}^{2}x_{34}^{2}%
}\text{+crossings}%
\end{equation}
into twist $2$ conformal blocks, one can find all the $p_{\Delta,l}$
coefficients. The $S$ and $T$ contributions can be disentangled since they
have a different index structure, see Eq.~(\ref{eq:son-s}). Even though we
will not need the explicit expressions, we quote the result:%
\begin{gather}
p_{l+2,l}^{T}=(l!)^{2}/(2l)!,\qquad p_{l+2,l}^{S}=(2/N)p_{l+2,l}^{T}%
\qquad(l\text{ even})\,,\nonumber\\
p_{l+2,l}^{A}=(l!)^{2}/(2l)!\qquad(l\text{ odd)}\,. \label{eq:freescalar}%
\end{gather}
Notice that all $p_{\Delta,l}\geq0$, consistent with unitarity.

We now proceed to showing that $any$ solution of the sum rule at $d_{\phi}=1$
must contain a singlet scalar of dimension $\leq2$. In fact we will show an
even stronger result---that any such solution must contain a singlet scalar of
dimension exactly $2$.

Let's group the twist $2$ terms in (\ref{eq:son-vect}) separately from the
rest:%
\begin{equation}
\sum_{\text{twist \thinspace2}}p_{\alpha}\mathbf{x}_{\alpha}+\sum_{\text{twist
}\neq2}p_{\beta}\mathbf{x}_{\beta}=\mathbf{y} \label{eq:separate}%
\end{equation}
First we will show that in any solution all twist $\neq2$ coefficients
$p_{\beta}$ must be zero. This is shown by exhibiting a linear functional
$\Lambda_{0}:V\rightarrow\mathds{R}$ such that%
\begin{align}
\Lambda_{0}[\mathbf{y}]  &  =0,\quad\nonumber\\
\Lambda_{0}[\mathbf{x}_{\alpha}]  &  =0\quad\forall\text{ fields of twist
}2\text{,}\label{eq:lambda0}\\
\Lambda_{0}[\mathbf{x}_{\beta}]  &  >0\quad\forall\text{ fields of twist }%
\neq2.\nonumber
\end{align}
Applying $\Lambda_{0}$ to (\ref{eq:separate}), and using the fact that
$p_{\beta}$ $\geq0$, we conclude immediately that all $p_{\beta}=0$.

The functional $\Lambda_{0}$ can be written explicitly by Taylor-expanding the
vectors entering the sum rule around the point $z=\bar{z}=1/2$. Let's
introduce the coordinates $a,b$%
\begin{equation}
z=\frac{1}{2}+a+b,\quad\bar{z}=\frac{1}{2}+a-b\,.
\end{equation}
Then the functions $F_{\Delta,l}$ and $H_{\Delta,l}$ are even with respect to
both $a$ and $b$, so that their derivatives $\partial_{a}^{m}\partial_{b}^{n}$
at $a=b=0$ are nonzero only if both $m$ and $n$ are even. It turns out that
the functional $\Lambda_{0}$ can be chosen as the following linear combination
of second derivatives at the $a=b=0$ point:%
\begin{equation}
\Lambda_{0}\left[  \mathbf{x}\right]  =A(\partial_{a}^{2}x_{1}-\partial
_{b}^{2}x_{1})+B(\partial_{a}^{2}x_{2}-\partial_{b}^{2}x_{2})+C(\partial
_{a}^{2}x_{3}-\partial_{b}^{2}x_{3})\,.
\end{equation}
Here $x_{1,2,3}$ are the components of the vector-function: $\mathbf{x}%
=(x_{1},x_{2},x_{3})^{T}\,.$ There is a certain freedom in choosing the
coefficients $A,B,C$; the following choice is one possibility which works for
all $N\geq2$:%
\begin{equation}
A=1,\,B=2,\,C=0\,.
\end{equation}
The first relation (\ref{eq:lambda0}) is trivially satisfied; the other two
follow from the following curious property of the $F_{\Delta,l}$ second
derivatives at the $a=b=0$ point:%
\begin{align}
\partial_{a}^{2}F_{\Delta,l}  &  =\partial_{b}^{2}F_{\Delta,l}\text{\qquad at
twist }2\,,\nonumber\\
\partial_{a}^{2}F_{\Delta,l}  &  >\partial_{b}^{2}F_{\Delta,l}\text{\qquad at
twist }\neq2\,.
\end{align}

Since all $p_{\beta}=0$, we are reduced to a simpler equation which involves
only twist $2$ fields:%
\begin{equation}
\sum_{\text{twist \thinspace2}}p_{\alpha}\mathbf{x}_{\alpha}=\mathbf{y}\,.
\end{equation}
We want to show that if we drop the scalar singlet from this equation, there
are no solutions. This is shown by exhibiting a second linear functional
$\Lambda_{1}$ with the following properties:%
\begin{align}
\Lambda_{1}[\mathbf{y}]  &  <0\,,\nonumber\\
\Lambda_{1}[\mathbf{x}_{\alpha}]  &  \geq0\text{ on all twist }2\text{ fields
except for the singlet scalar.}%
\end{align}
Notice that any such $\Lambda_{1}$ must necessarily be negative on the singlet
scalar, to allow at the very least the existence of one explicit solution
(\ref{eq:freescalar}).

This functional can be written again as a linear combination of derivatives at
$a=b=0$:%
\begin{equation}
\Lambda_{1}[\mathbf{x}]=\sum_{i=1}^{3}\sum_{\substack{m,n\text{ even}\\0\leq
m+n\leq k}}\frac{\lambda_{m,n}^{i}}{m!n!}\partial_{a}^{m}\partial_{b}^{n}%
x_{i}\,. \label{eq:lambdader}%
\end{equation}
At present we can only find the coefficients $\lambda_{m,n}^{i}$ numerically.
For $2\leq N\leq7$ it is enough to use derivatives up to the second order,
just like in $\Lambda_{0}$. For example, for $N=4$ one can use the functional
whose only nonzero coefficients are%
\begin{equation}%
\begin{array}
[c]{ll}%
\lambda_{0,0}^{1}=-8\,, & \lambda_{0,2}^{1}=6\,,\\
\lambda_{0,0}^{2}=-11\,, & \lambda_{0,2}^{2}=8\,,\\
\lambda_{0,0}^{3}=-10\,, & \lambda_{0,2}^{3}=3\,.
\end{array}
\,
\end{equation}
Including derivatives up to the fourth order $(k=4$) allows to find
functionals in the range up to $N\leq128$. While we have not checked higher
$N$, we feel sufficiently confident that, adding more and more derivatives,
functional $\Lambda_{1}$ can be found for any $N$. With this small proviso,
the demonstration that at $d_{\phi}=1$ there must be a dimension $2$ singlet
scalar is complete.

\begin{figure}
[h]
\begin{center}
\includegraphics[
height=2.2441in,
width=2.6964in
]%
{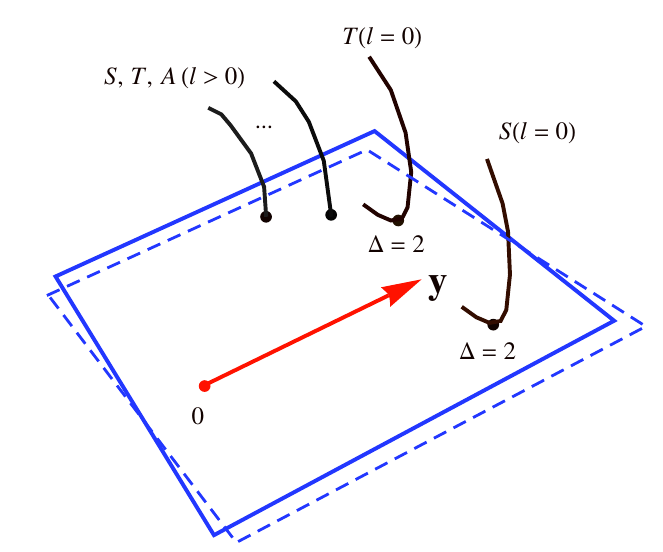}%
\caption{\textit{This figure gives a geometric interpretation of the proof
that at }$d_{\phi}=1$\textit{ the sum rule has no solution unless the }%
$\Delta=2$\textit{ singlet scalar is included in the spectrum. The
solid-contour plane represents the zero set of the functional }$\Lambda_{0}%
$\textit{. The vector }$\mathbf{y}$\textit{ and all the twist }$2$\textit{
vectors (black dots) lie in the }$\Lambda_{0}=0$\textit{ plane. On the other
hand, all twist }$\neq2$\textit{ vectors, which for varying }$\Delta$\textit{
trace separate curves labeled by spin and representation, lie on one side of
this plane (}$\Lambda_{0}>0)$\textit{. The }$\Lambda_{0}=0$\textit{ plane can
be slightly rotated so that the }$\mathbf{y}$\textit{ vector and the twist 2
singlet scalar lie on one side of the rotated plane, while the rest of the
twist }$2$\textit{ vectors lie on the opposite side. The rotated plane (dashed
contour) can be described by an equation }$\Lambda_{0}+\varepsilon\Lambda
_{1}=0$\textit{ for a small }$\varepsilon$\textit{.}}%
\label{fig:d=1}%
\end{center}
\end{figure}

It is perhaps useful to give a geometric representation of the given proof,
see Fig.~\ref{fig:d=1}. The existence of $\Lambda_{0}$ means that there is a
hyperplane (the zero set of $\Lambda_{0}$) such that all the twist 2 vectors
as well as the vector $\mathbf{y}$ belong to it, while all twist $\neq2$
vectors lie on one side of it. The existence of $\Lambda_{1}$ means that this
hyperplane can be rotated so that the twist 2 singlet scalar and the rest of
the twist $2$ vectors lie on the opposite sides of the rotated hyperplane.%

Let us now discuss the $SU(N)$ case. The statement is the same: at $d_{\phi
}=1$ the OPE $\phi\times\phi^{\dagger}$ must contain a dimension $2$ singlet
scalar. This is derived from the $SU(N)$ vectorial sum rule (\ref{eq:sun-rule}%
) by using the same method of $\Lambda_{0}$ and $\Lambda_{1}$. Just like for
$SO(N),$ the functional $\Lambda_{0}$ can be given by using only second
derivatives. For $\Lambda_{1}$, one can use derivatives up to the second order
for $N=2,3$, while derivatives up to the fourth order work for at least all
$N\leq50$.

Finally, the reader may want to compare the above discussion with Section 5.4
of \cite{r1}, where the $d_{\phi}=1$ case was analyzed analogously, and
somewhat more explicitly, for CFTs without global symmetry.

\subsection{Some numerical results at $d_{\phi}>1$}

As explained in Section \ref{sec:generalities}, the bound $f_{S}$ is a
continuous function of $d_{\phi}.$ The previous Section has established that
$f_{S}=2$ at $d_{\phi}=1$. The next question is to understand how fast the
bound deviates from $2$ at we increase $d_{\phi}$ above $1$. To do this we
need to be able to compute the bound at any given $d_{\phi}$. This is done by
the method of linear functionals \cite{r1}, already used above to analyze the
$d_{\phi}=1$ case. At any fixed $d_{\phi}$, we will be looking for a linear
functional $\Lambda$ such that%
\begin{align}
\Lambda\lbrack\mathbf{y}]<0,  &  \quad\nonumber\\
\Lambda\lbrack\mathbf{x}_{\alpha}]\geq0  &  \qquad\forall\text{ scalar
singlets with }\Delta\geq\Delta_{S}\text{ and}\label{eq:lambda3}\\
&  \qquad\forall\text{ other fields (subject to the unitarity bounds)\,.}%
\nonumber
\end{align}
If such a functional exists, then the spectrum without any singlet scalars
below $\Delta_{S}$ cannot solve the vectorial sum rule and cannot be realized
in any CFT. Geometrically, Eqs.~(\ref{eq:lambda3}) says that there is a
hyperplane separating the $\mathbf{y}$ vector from the cone generated by such
a spectrum. The bound $f_{S}$ is computed as the smallest $\Delta_{S}$ for
which a functional satisfying (\ref{eq:lambda3}) can be found.

Assuming that $\Lambda$ is of the form (\ref{eq:lambdader}),
Eqs.~(\ref{eq:lambda3}) define a linear programming problem for the
coefficients $\lambda_{m,n}^{i}$. However, the number of constraints is
formally infinite, which requires careful discretizations and truncations. We
refer to \cite{r1},\cite{r2}, and also to \cite{poland}, for a detailed
description of how these numerical difficulties are overcome.

There is no difference of principle between the system (\ref{eq:lambda3}) and
the analogous system analyzed in the case without global symmetry in our
previous work; it is just bigger. There are several families of vectors
$\mathbf{x}_{\alpha}$, labeled by representation and spin parity, and each
vector now represents a \textit{vector}-function. As a result, for the same
number of derivatives $($parameter $k$ in (\ref{eq:lambdader})), the globally
symmetric case involves $Q$ times more constraints and $Q$ times more
functional coefficients, where $Q=3;6$ for $SO(N);SU(N).$ This makes
computations much more time-consuming as well as increases numerical
instabilities. In the case without global symmetry, the initial analysis of
\cite{r1} used $k=6$, and subsequently\ we were able to push $k$ up to $18$
\cite{r2}, producing a very strong bound. For the globally symmetric case, we
have so far not been able to go beyond $k=4$ for $SO(N)$ and $k=2$ for $SU(N)$.

Here is an account of these exploratory calculations:

\begin{itemize}
\item In every case that we looked at, we found a bound for $d_{\phi}$ near
$1$ which approached $2$ continuously in the $d_{\phi}\rightarrow1$ limit.
Thus we have checked the existence theorem from the previous Section.

\item We have seen that the bound is a monotonically increasing function of
$d_{\phi}$ at $d_{\phi}>1$.

\item Typically, we have only computed the bound in the interval $1\leq
d_{\phi}\leq d_{\ast}$ where $f_{S}(d_{\ast})\approx4.$ This is because
knowing whether the singlet scalar is relevant or irrelevant is a particularly
interesting question. It is also sufficient to get an idea about how strong
the bound is, for a given $k$. In the following table, we give the $d_{\ast}$
for the cases that we considered. At the current level of accuracy, our
$f_{S}$ interpolates almost linearly between $2$ and $4$ as $d_{\phi}$
increases from $1$ to $d_{\ast}$.
\end{itemize}

\begin{table}[h]
\begin{center}%
\begin{tabular}
[c]{|c||c|c|c||c|c|}\hline
$G$ & $U(1)\equiv SO(2)$ & $SO(3)$ & $SO(4)$ & $SU(2)$ & $SU(3)$\\\hline
$d_{\ast}$ &
\begin{tabular}
[c]{c}%
$1.063$ $(k=2)$\\
$1.12~(k=4)$%
\end{tabular}
&
\begin{tabular}
[c]{c}%
$1.032~(k=2)$\\
$1.08~(k=4)$%
\end{tabular}
&
\begin{tabular}
[c]{c}%
$1.017~(k=2)$\\
$1.06~(k=4)$%
\end{tabular}
&
\begin{tabular}
[c]{c}%
$1.016$\\[-5pt]%
$(k=2)$%
\end{tabular}
&
\begin{tabular}
[c]{c}%
$1.003$\\[-5pt]%
$(k=2)$%
\end{tabular}
\\\hline
\end{tabular}
\end{center}
\caption{\textit{First line: the global symmetry groups considered. The
external scalar was assumed to transform in the fundamental. Second line: the
value of external scalar's dimension }$d_{\phi}$\textit{ for which the bound
}$f_{S}$\textit{ on the singlet scalar dimension (monotonically increasing
from }$2$\textit{ for }$d_{\phi}=1$\textit{) was seen to cross }$4.$}%
\label{tab:1}%
\end{table}

As we already mentioned in Section \ref{sec:U(1)}, a $U(1)$ bound on the
singlet scalar dimension was recently published in \cite{poland}. It is not
possible to compare our and their results directly: on the one hand, they
assume supersymmetry and expand to a very high order $(k=12$) which should
make their bound stronger, but on the other hand they use but one scalar
component of the full vectorial sum rule, which makes their bound weaker. It
seems that these two effects compensate each other, so that their bound is
roughly comparable to our $U(1),$ $k=4$ bound. Notice however that they could
not see any bound for $d\geq1.16$, while we checked by using the full
vectorial sum rule that a bound continues to exist even for larger $d$.

\section{Discussion}

In this paper we extended the constraints from OPE associativity and
crossing first derived in Ref.~\cite{r1} to CFTs with global
symmetry. Focussing on the scalar 4-point function we derived a set
of sum rules that constrains the operator content in all the
possible channels with given symmetry and parity $(-1)^{\ell}$
quantum numbers. More precisely, by explicit examples and by a
general argument, we have shown that the number of sum rules equals
the number of possible channels. This results suggest that, in
principle, one could obtain (correlated) constraints on the
operators appearing in each channel. In analogy with previous
studies these constraints could involve the dimension and the fusion
coefficients of the lowest lying operators in a given channel.

As a first exploration we have studied the possibility of obtaining
an upper bound on the dimension of the scalar singlet of lowest
dimension appearing in the OPE of $\phi\times\phi$ and
$\phi\times\phi^{\dagger}$. That question is also relevant to asses
the viability of the so called Conformal Technicolor (CT) scenario
\cite{luty}. The goal of CT is to achieve a natural separation
between the electroweak scale and the scale of Flavor. That is
motivated by the experimental success of the
Cabibbo-Kobayashi-Maskawa pattern of flavor violation. The way CT
works is that the electroweak symmetry breaking sector above the
weak scale flows to a strongly coupled CFT. The role of the Higgs
field is then played by a composite operator $H$, while electroweak
gauge interactions arise from the weak gauging of a $SU(2)\times
U(1)$ subgroup of the global symmetry group $G$ of the CFT. In that
situation, suppression of flavor violation is the more robust the
closer to $1$ is the dimension $d_{H}$ of $H$. On the other hand, a
natural separation of mass scale requires all total singlet scalars
to be only marginally relevant or irrelevant. Indicating by $S$ the
lowest dimension singlet $\subset H\times H^{\dagger}$, we thus need
$d_{S}$ slightly below or above $4$ \footnote{Of course if $d_{S}$
is strictly $>4$ we need some other marginally relevant coupling, or
a strongly relevant coupling which can be taken small because of an
extra symmetry, to generate the weak scale by dimensional
transmutation.}. A major constraint on
$d_{H}$ is imposed by the top coupling, that runs like $\lambda_{t}%
(\mu)=\lambda_{t}(\mu_{EW})(\mu/\mu_{EW})^{d_{H}-1}$ and quickly becomes
strong for $d_{H}-1=O(1)$. Notice that one major difficulty to achieve the
dream of CT is that for $d_{H}=1$ we must have $d_{S}=2$ and the hierarchy
problem in all its splendor.

To be more quantitative about the needed pattern of field dimensions,
assumptions on the physics of flavor must be made. Making the optimistic,
although plausible, assumption that flavor violation in the light families is
either suppressed by their mixing to the third family of by their Yukawa
couplings \footnote{For instance for the $\Delta S=2$ operators contributing
to $K\bar{K}$-mixing, this amounts to assuming respectively an extra
suppression factor $\sim(V_{st}V_{dt})^{2}$ or $y_{s}y_{d}/y_{t}^{2}$ compared
to operators involving just the third family quarks.}, the range
$d_{H}\mathrel{\vcenter {\hbox{$<$}\nointerlineskip\hbox{$\sim$}}}1.7$,
$d_{S}\mathrel{\vcenter
{\hbox{$>$}\nointerlineskip\hbox{$\sim$}}}3.5\div4$ is sufficient. In that
situation the scale where the top Yukawa becomes strong can be as low as
$\sim100$ TeV, so that the window where CT is active is not very big. On the
other hand, the more conservative, but robust, assumption that all flavor
violating operators are equally important at the Flavor scale requires the
more constrained pattern $d_{H}-1\mathrel{\vcenter
{\hbox{$<$}\nointerlineskip\hbox{$\sim$}}}0.2$, $d_{S}\geq4$. That second
situation corresponds to a flavor scale around $10^{5}$ TeV, with the CFT
describing physics in a sizeable window of scales. In view of the above, it
would be interesting to derive, if it exists, an upper bound on $d_{S}$ as a
function of $d_{H}$ which was one goal of the present paper. Indeed in
\cite{r1} a first step toward an answer was given by deriving an upper bound
on $\mathrm{min}(d_{S},d_{T})$, where $d_{T}$ is the dimension of the triplet
scalar $\subset H\times H^{\dagger}$. The method of \cite{r1} was however
\textquotedblleft symmetry blind" in that it could not resolving the singlet
and triplet channels. In \cite{r2} a further refinement of the bound was
obtained by working up to 18 derivatives in function space. The bound is a
monotonically growing function of $d_{H}$ crossing 4 at about $d_{H}=1.6$.
That result is compatible with the flavor-optimistic CT scenario. Given the
clear signs of convergence of the symmetry blind bound \cite{r2}, and also
given the remarkable success of the method in 2D CFTs where the bound
basically tracks minimal models, we are tempted to conclude that the
flavor-optimistic CT scenario is plausible. However the bound of \cite{r2}, if
interpreted as a bound on $d_{S}$ (that is if $d_{S}<d_{T}$), would thoroughly
rule out the flavor-robust CT.

In the present paper we have instead shown that it is possible to obtain an
independent bound on the singlet. We have rigorously shown that the bound on
$d_{S}$ exists and goes smoothly to 2 as $d_{\phi}\rightarrow1$ for $\phi$ a
fundamental in $SO(N)$ ($SU(N)$) and $N\leq128$ ($N\leq50$). We have further
worked it out numerically for a few small groups, and in particular for
$SO(4)$, which is the smallest group of phenomenological relevance. The
results are not yet very strong, as seen in Table 1. In particular, in
$SO(4)$, $d_{S}$ crosses 4 already for $d_{\phi}=1.06$, way within the
interesting region of flavor-robust CT. One reason for the weakness of the
bound is that our numerical method based on the Linear Programming algorithm
does not converge fast enough when the function space is truncated beyond
$k=4$ derivatives. So our best bound just corresponds to working up to $k=4$.
One reason of the extra difficulty with respect to the symmetry blind case is
that we are now dealing with a triple sum rule, rather than with a single one,
and the complexity grows $3^{2}=9$ times faster with $k$. Notice that even in
the symmetry blind case the bounds at low $k$ are not very strong. Indeed one
has the progression $d_{\ast}\approx1.12\ (k=2)$ \cite{r1}, $d_{\ast}%
\approx1.18$ $(k=4)$ (unpublished), $d_{\ast}\approx1.35$ ($k=6)$ \cite{r1},
eventually increased to $d_{\ast}\approx1.6$ for $k=18$ \cite{r2}. Assuming a
similar rate of improvement for the globally symmetric bound, and assuming
optimistically that we could push the analysis to similarly high values of $k$
(which would likely require new ideas in algorithm implementation), we could
expect to get to $d_{\ast}\approx$ $1.1\div1.2$ for the $SO(4)$ case. This is
more or less at the edge of interest of the flavor-robust CT.

It is important to understand why the globally symmetric bounds are
so weak, and why they are getting even weaker at larger $N$, as
shown in Table 1. One could imagine two alternative explanations.
One, boring, possibility is that crossing symmetry for the 4-point
function of just one operator $\phi_{\alpha}$ is simply not an
efficient constraint in presence of global symmetry. It would be
more interesting if, perhaps, our result is telling us something
physical, namely that for larger global symmetry, the role of the
singlet $S$ in maintaining consistency of the theory is indeed
getting smaller, so that it can be allowed to decouple. A very
partial hint of that could be the fact, already emphasized in
\cite{r1}, that in the $O(N)$ model in $4-\epsilon$ dimensions the
anomalous dimension of the singlet is $O(\epsilon)$ while that of
the symmetric traceless is $O(\epsilon/N)$. One way to test which of
the two possibilities is true is to derive a twin bound on the
symmetric traceless $T$ (without imposing any constraints on the
singlet $S$). If also that bound were found to be weak, the first
possibility would be favored. If, on the other hand, the symmetric
traceless bound would turn up much stronger than the singlet one,
and perhaps comparable in strength
to the general bound of \cite{r1}, \cite{r2} on the $\mathrm{min}(d_{T}%
,d_{S})$, then this would be an indication that tensor is much more
important than singlet in maintaining the crossing symmetry. Note
that, as far as we know, there's no simple reason why the bounds get
weaker with larger $N$. At this stage this is just an experimental
fact.

There is yet another piece of information which, if taken into account, could
change the picture qualitatively. In our numerical study we did not make any
assumption about the stress tensor and symmetry current central charges. In
principle our bound could become stronger under the condition that these
central charges are bounded from above, corresponding to a perhaps more
reasonable theory, that is one that does not contain too many degrees of
freedom. Indeed, in connection to our main phenomenological application, the
common wisdom is that constraints from the $S$ parameter point towards a small
EWSB sector, so that the central charges should be small. This is especially
true of the current central charge, because the current-current spectral
density enters directly into the spectral representation of the $S$ parameter.
Since conformal symmetry is broken in the IR, only the high energy tail of
this density will be controlled by the CFT central charge. This does not allow
to make this connection precise, but still large central charge seems to be disfavored.

What one could do then is to study a lower bound on the central charge as a
function of the gap in the singlet scalar sector. For the stress tensor
central charge in the case without global symmetry, precisely such a study was
performed in \cite{r3}. These studies are made possible by the existence of
$O(1)$ universal bounds on the coefficients $p_{\Delta,l}$ in the conformal
block decomposition \cite{cr}, and that the coefficients $p_{4,2}$ and
$p_{3,1}$ can be related to the inverse central charges by the Ward identities
\cite{op}. In \cite{r3}, we found that higher gap sometimes requires a
significant increase in the stress tensor central charge. If the current
central charge is shown to have an even stronger dependence on the singlet
scalar gap, this may indicate a potential difficulty with the $S$ parameter
for CT.

 Notice that central charge studies are also
interesting in their own right, without connection to CT. For
example, can one show that an $SO(N)$ theory with a fundamental
necessarily has central charges larger than free theory of $N$
scalars? For the stress tensor central charge and the case without
global symmetry this was shown in \cite{r3},\cite{poland} (in a
range of $d_{\phi}$ near $1$).

In passing, one could imagine one day bounds like those discussed in
this paper would make contact with the studies of IR fixed points of
gauge theories performed on the lattice. Since the very existence of
an IR fixed point implies that these theories cannot contain a
singlet scalar with dimension below 4, our bounds could provide
rigorous theoretical constraints on the lattice measurements of the
fermion bilinear operator dimension \cite{lattice}.

The framework laid out in this paper will likely lead to many
applications beyond those mentioned above. Be aware that the next
crucial steps are an algorithm improvement and/or finding if there
exists a set of questions for which the convergence is faster, so
that interesting bounds can be obtained already at small $k$. This
is a new field to explore!

\subsection*{Acknowledgements}

We are grateful to Erik Tonni for collaboration at the early stages of this
project. The work of R.R. and A.V. is supported by the Swiss National Science
Foundation under contract No. 200020-126941. The work of S.R. was supported in
part by the European Programme \textquotedblleft Unification in the LHC Era",
contract PITN-GA-2009-237920 (UNILHC).


\begin{thebibliography}{99}                                                                                               %


\bibitem {r1}{ R.~Rattazzi, V.~S.~Rychkov, E.~Tonni and A.~Vichi,
\textquotedblleft Bounding scalar operator dimensions in 4D
CFT,\textquotedblright\ JHEP \textbf{0812}, 031 (2008)
\href{http://arxiv.org/abs/0807.0004}{arXiv:0807.0004}.
%%CITATION = JHEPA,0812,031;%%
}

\bibitem {r2}V.~S.~Rychkov and A.~Vichi, \textquotedblleft Universal
Constraints on Conformal Operator Dimensions,\textquotedblright%
\ Phys.\ Rev.\ D \textbf{80}, 045006 (2009)
\href{http://arxiv.org/abs/0905.2211}{arXiv:0905.2211}.
%%CITATION = PHRVA,D80,045006;%%


\bibitem {cr}F.~Caracciolo and V.~S.~Rychkov, \textquotedblleft Rigorous
Limits on the Interaction Strength in Quantum Field Theory,\textquotedblright%
\ Phys.\ Rev.\ D \textbf{81}, 085037 (2010)
\href{http://arxiv.org/abs/0912.2726}{arXiv:0912.2726}.
%%CITATION = PHRVA,D81,085037;%%


\bibitem {r3}R.~Rattazzi, S.~Rychkov and A.~Vichi, ``Central Charge Bounds in
4D Conformal Field Theory,'' arXiv:1009.2725 [hep-th].
%%CITATION = ARXIV:1009.2725;%%


\bibitem {pol}A.~M.~Polyakov, \textquotedblleft Nonhamiltonian approach to
conformal quantum field theory,\textquotedblright%
\ Zh.\ Eksp.\ Teor.\ Fiz.\ \textbf{66}, 23 (1974).
%%CITATION = ZETFA,66,23;%%


\bibitem {poland}D.~Poland and D.~Simmons-Duffin, \textquotedblleft Bounds on
4D Conformal and Superconformal Field Theories,\textquotedblright%
\ \href{http://arxiv.org/abs/1009.2087}{arXiv:1009.2087} [hep-th].
%%CITATION = ARXIV:1009.2087;%%


\bibitem {luty}{ M.~A.~Luty and T.~Okui, \textquotedblleft Conformal
technicolor,\textquotedblright\ JHEP \textbf{0609}, 070 (2006)
\href{http://arxiv.org/abs/hep-ph/0409274}{arXiv:hep-ph/0409274}.
%%CITATION = JHEPA,0609,070;%%
M.~A.~Luty, \textquotedblleft Strong Conformal Dynamics at the LHC
and on the Lattice,\textquotedblright%
\ \href{http://arxiv.org/abs/0806.1235}{arXiv:0806.1235}.
%%CITATION = ARXIV:0806.1235;%%
}J.~Galloway, J.~A.~Evans, M.~A.~Luty and R.~A.~Tacchi, ``Minimal Conformal
Technicolor and Precision Electroweak Tests,''
\href{http://arxiv.org/abs/1001.1361}{arXiv:1001.1361} [hep-ph].
%%CITATION = ARXIV:1001.1361;%%


\bibitem {do12}{ F.~A.~Dolan and H.~Osborn, \textquotedblleft Conformal four
point functions and the operator product expansion,\textquotedblright%
\ Nucl.\ Phys.\ B \textbf{599}, 459 (2001)
\href{http://arxiv.org/abs/hep-th/0011040}{arXiv:hep-th/0011040}.
%%CITATION = NUPHA,B599,459;%%
\textquotedblleft Conformal partial waves and the operator product
expansion,\textquotedblright\ Nucl.\ Phys.\ B \textbf{678}, 491 (2004)
\href{http://arxiv.org/abs/hep-th/0309180}{arXiv:hep-th/0309180}.
%%CITATION = NUPHA,B678,491;%%
}

\bibitem {slansky}R.~Slansky,
%``Group Theory For Unified Model Building,''
Phys.\ Rept.\ \textbf{79}, 1 (1981).
%%CITATION = PRPLC,79,1;%%


\bibitem {mack}S.~Ferrara, R.~Gatto and A.~F.~Grillo, \textquotedblleft
Positivity Restrictions On Anomalous Dimensions,\textquotedblright%
\ Phys.\ Rev.\ D \textbf{9}, 3564 (1974);
%%CITATION = PHRVA,D9,3564;%%
{G.~Mack, \textquotedblleft All Unitary Ray Representations Of The Conformal
Group SU(2,2) With Positive Energy,\textquotedblright%
\ \href{http://projecteuclid.org/DPubS?service=UI&version=1.0&verb=Display&handle=euclid.cmp/1103900926}{Commun.\ Math.\ Phys.\ \textbf{55}%
, 1 (1977)}.
%%CITATION = CMPHA,55,1;%%
}

\bibitem {op}H.~Osborn and A.~C.~Petkou, \textquotedblleft Implications of
Conformal Invariance in Field Theories for General
Dimensions,\textquotedblright\ Annals Phys.\ \textbf{231}, 311 (1994)
\href{http://arxiv.org/abs/hep-th/9307010}{arXiv:hep-th/9307010}.
%%CITATION = APNYA,231,311;%%


\bibitem {lattice}F.~Bursa, L.~Del Debbio, L.~Keegan, C.~Pica and T.~Pickup,
``Mass anomalous dimension in SU(2) with two adjoint fermions,''
Phys.\ Rev.\ D \textbf{81}, 014505 (2010)
\href{http://arxiv.org/abs/0910.4535}{arXiv:0910.4535} [hep-ph].
%%CITATION = PHRVA,D81,014505;%%
L.~Del Debbio, B.~Lucini, A.~Patella, C.~Pica and A.~Rago, ``The infrared
dynamics of Minimal Walking Technicolor,'' Phys.\ Rev.\ D \textbf{82}, 014510
(2010) \href{http://arxiv.org/abs/1004.3206}{arXiv:1004.3206} [hep-lat].
%%CITATION = PHRVA,D82,014510;%%
T.~DeGrand, Y.~Shamir and B.~Svetitsky, ``Running coupling and mass anomalous
dimension of SU(3) gauge theory with two flavors of symmetric-representation
fermions,'' \href{http://arxiv.org/abs/1006.0707}{arXiv:1006.0707} [hep-lat].
%%CITATION = ARXIV:1006.0707;%%
F.~Bursa, L.~Del Debbio, L.~Keegan, C.~Pica and T.~Pickup, ``Mass anomalous
dimension in SU(2) with six fundamental fermions,''
\href{http://arxiv.org/abs/1007.3067}{arXiv:1007.3067} [hep-ph].
%%CITATION = ARXIV:1007.3067;%%

\end{thebibliography}
\end{document}